\documentclass{jfm}
\usepackage{graphicx}
\usepackage[dvipsnames]{xcolor}
\shorttitle{Experimental study of the non-linear saturation of the elliptical instability}
\shortauthor{T. Le Reun, B. Favier and M. Le Bars}

\title{Experimental study of the non-linear saturation of the elliptical instability: inertial wave turbulence versus geostrophic turbulence}

\author{Thomas Le Reun \aff{1}
  \corresp{\email{lereun@irphe.univ-mrs.fr}},
  Benjamin Favier \aff{1}
 \and Michael Le Bars\aff{1}}

\affiliation{\aff{1} Aix Marseille Univ, CNRS, Centrale Marseille, IRPHE UMR 7342, Marseille, France
}

\xdefinecolor{RED}{named}{black}

\newcommand{\mbf}{\boldsymbol}

\newcommand{\ie}{\textit{i.e.}~}
\newcommand{\bu}{\mbf{u}}
\newcommand{\bU}{\mbf{U}}
\newcommand{\bX}{\mbf{X}}
\newcommand{\bx}{\mbf{x}}
\newcommand{\der}[2]{\frac{\mathrm{d} #1}{\mathrm{d} #2}}
\newcommand{\BF}{\mbf{U}_{b}^{\Omega}}
\newcommand{\BFL}{\mbf{U}_{b}^{\ell}}
\newcommand{\MF}{\overline{\bU}}

\begin{document}

\maketitle

\begin{abstract}
In this paper, we present an experimental investigation of the turbulent saturation of the flow driven by parametric resonance of inertial waves in a rotating fluid.
In our set-up, a half-meter wide ellipsoid filled with water is brought to solid body rotation, and then undergoes sustained harmonic modulation of its rotation rate.
This triggers the exponential growth of a pair of inertial waves via a mechanism called the libration-driven elliptical instability. 
Once the saturation of this instability is reached, we observe a turbulent state for which energy is injected into the resonant inertial waves only. 
Depending on the amplitude of the rotation rate modulation, two different saturation states are observed.
At large forcing amplitudes, the saturation flow mainly consists of a steady, geostrophic anticyclone.
Its amplitude vanishes as the forcing amplitude is decreased while remaining above the threshold of the elliptical instability.
Below this secondary transition, the saturation flow is a superposition of inertial waves which are in weakly non-linear resonant interaction, a state that could asymptotically lead to inertial wave turbulence.
In addition to being a first experimental observation of a wave-dominated saturation in unstable rotating flows, the present study is also an experimental confirmation of the model of \cite{le_reun_inertial_2017-1} who introduced the possibility of these two turbulent regimes.
The transition between these two regimes and their relevance to geophysical applications are finally discussed.

\end{abstract}

\section{Introduction}

Tides are fundamental interactions between astrophysical bodies that cause deformation of their shapes and alteration of their rotation rates.
The fluid interiors of these bodies, be it a surface or a sub-surface ocean, a liquid metallic planetary cores or a stellar interior, are known to be sensitive to tidal excitation which drives a wealth of flows.
Tides may play a role in the orbital evolution of planets and stars \citep{ogilvie_tidal_2004-1,goodman_dynamical_2009,le_bars_tidal_2010,barker_non-linear_2016}, and in dynamo action and magnetic field generation inside terrestrial planets \citep{
kerswell_tidal_1998,
le_bars_impact-driven_2011,dwyer_long-lived_2011,
cebron_elliptical_2012,
le_bars_flows_2015-1,vidal_magnetic_2018}.
The tidal excitation of flows is primarily driven by the shape deformation of fluid interiors, combined with non-stationary effects.
The latter include differential rotation between the astrophysical body and its tidal bulge (as is the case of the Earth or any body whose spin is not synchronised with its companion), libration (periodic oscillations of the rotation rate), but also precession and nutation (periodic variations of the rotation axis direction).
%
%
%
These harmonic mechanical forcings can directly excite inertial waves \citep{aldridge_axisymmetric_1969,goodman_dynamical_2009}, drive zonal flows \citep{morize_experimental_2010-1,sauret_tide-driven_2014,favier_non-linear_2014}, and generate turbulence from various instabilities. 
Among them is the elliptical instability, which is the parametric, subharmonic resonance of a pair of inertial waves with the tidal base flow  \citep{kerswell_elliptical_2002,le_bars_flows_2015-1}.
It has been shown to be caused by either the differential rotation of the tidal bulge and the planet \citep{malkus_experimental_1989-1,
le_dizes_three-dimensional_2000,
le_bars_coriolis_2007,
le_bars_tidal_2010} or by libration \citep{kerswell_tidal_1998,cebron_libration_2012,cebron_libration-driven_2014}.
This is particularly interesting in the case of planetary cores as it provides an alternative stirring mechanism to thermal and solutal convection usually invoked to explain dynamo action, a phenomenon that has been described in particular by \cite{reddy_turbulent_2018} in their study of kinematic dynamos driven by tidal instabilities.
Although the mechanism of the elliptical instability is very well understood, its non-linear saturation exhibits a variety of behaviours. 
On the one hand, several experimental \citep{malkus_experimental_1989-1,eloy_experimental_2000} and numerical \citep{barker_non-linear_2013-1,barker_non-linear_2016-1}  works have reported that the instability gives rise to cycles of inertial wave exponential growth alternating with turbulent break-downs. 
As the instability reaches its saturation, \cite{barker_non-linear_2013-1} have in particular shown that strong geostrophic vortices, that is, vortices invariant along the rotation axis of the fluid, emerge and disrupt the instability mechanism.
On the other hand, other set-ups of the elliptical instability produce sustained turbulence \citep{grannan_experimental_2014,favier_generation_2015}, in particular in the presence of a weak background magnetic field \citep{barker_non-linear_2014}. 
This clear dichotomy has been partially explained in the idealised study carried out by \cite{le_reun_inertial_2017-1} where the use of an adjustable, artificial friction specific to geostrophic flows allowed to continuously transition from one type of saturation to another.
The inhibition of the instability has been understood as a consequence of the resonant inertial wave frequency detuning induced by the geostrophic vortices as they take over the flow \citep{barker_non-linear_2013-1,le_reun_inertial_2017-1}.
Conversely, the sustained turbulent motion obtained by \cite{le_reun_inertial_2017-1} has been shown to be reminiscent of inertial wave turbulence, \ie a non-linear state made of a superposition of waves in non-linear resonant interaction \citep{galtier_weak_2003,bellet_wave_2006}.
This wave turbulence state has been reported in various other contexts such as surface capillary waves \citep{aubourg_nonlocal_2015}, vibrating plates \citep{during_weak_2006,miquel_nonstationary_2011} and internal waves in stratified fluids \citep{brouzet_energy_2016,le_reun_parametric_2018}.
Nevertheless, a clear experimental proof that it exists in rotating fluids undergoing mechanical forcing is still lacking, and the regime of parameters for which it could be observed remains to be determined.  
The duality between inertial waves and geostrophic vortices is also a defining feature of rotating turbulence.
Although geostrophic vortices are ubiquitously observed in experiments and simulations \citep{godeferd_structure_2015}, inertial waves have also been detected in various set-ups using sophisticated spectral techniques \citep{clark_di_leoni_quantification_2014,
yarom_experimental_2014,
yarom_experimental_2017,
oks_inverse_2017} or spatio-temporal correlations \citep{favier_space_2010,campagne_disentangling_2015}.
Several of these studies even quantify the detuning of the wave frequencies induced by the advection due to geostrophic vortices. 
However, classical experiments and numerical simulations of rotating turbulence are not adequate to fully understand the saturation of the elliptical instability: energy is usually injected in random structures or pair of vortices whereas the parametric resonance we are interested in only supplies energy to the fluid through inertial waves.
This instability is therefore a natural set-up to probe the non-linear fate of waves and the dependence of rotating turbulence with the forcing.
This paper presents an experimental study of the non-linear fate of the elliptical instability driven by libration.
It builds on the previous works of \cite{noir_experimental_2012-2} and \cite{grannan_experimental_2014}, although we explore more extreme regimes in which both the forcing amplitude and the dissipation are small.
The regimes attained here are more relevant for geo- and astrophysical and remain presently beyond the reach of global direct numerical simulations. 
We find at the lowest forcing amplitudes a regime which is dominated by inertial waves in non-linear resonant interaction, a state reminiscent of inertial wave turbulence. 
As the forcing amplitude is increased, a secondary transition associated with the emergence of a strong mean flow occurs; the saturation flow is then geostrophic-dominated instead of wave-dominated.
Our set-up allows to find the boundaries between these two regimes when the geophysically relevant control parameters are varied. 
The present article is organised as follows. 
A first section is devoted to introducing the forcing flow driven by librations as well as the basics of the elliptical instability and rotating turbulence.
We then describe the experimental set-up and the way through which the velocity fields are processed to investigate the flows driven by libration.
We finally detail our findings regarding the saturation of the instability, in particular the properties of the two regimes and their respective boundaries.

\section{Libration-driven flows and elliptical instability}

\subsection{Libration-driven flows}

Libration is a periodic modulation of the rotation rate of a body. 
We consider in our experiments the most simple form: 
\begin{equation}
\label{eq:libration}
\mbf{\Omega} = \Omega_0 \left( 1 + \varepsilon \sin\left(f \Omega_0 t \right) \right)  \mbf{e}_z
\end{equation}
where $\mbf{e}_z$ is a unit vector and $\varepsilon$ and $f$ are dimensionless parameters controlling respectively the amplitude and the frequency of the rotation rate modulation. 
When observed from a frame of reference rotating at $\mbf{\Omega}_0 = \Omega_0 \mbf{e}_z$, the solid outer boundary of the fluid periodically oscillates with an amplitude angle $\Delta \varphi = \varepsilon/f $. 

Librations drive an oscillating flow in tidally-distorted fluid interiors. 
To describe the flow driven by such a rotation rate modulation, let us consider an ellipsoid with principle axes of length $a$, $b$ and $c$, the third direction being parallel to $\boldsymbol{e}_z$. 
In the librating frame, \ie in the frame in which the outer boundary is fixed, consider a system of axes $(OXYZ)$ aligned with the principal axes of the ellipsoid. 
The inviscid libration-driven flow $\mbf{U}_b^{\ell}$, determined with the method of \cite{hough_xii._1895}, reads \citep{grannan_experimental_2014}:
\begin{equation}
\label{eq:libration_base_flow_librating_frame}
\mbf{U}_b^{\ell} = \left[
U ,
V ,
0
\right] = -\frac{2ab}{a^2 + b^2} \Omega_0 \varepsilon \sin(\Omega_0 f t)
\left[
-\displaystyle\frac{a}{b} Y ~,~
\displaystyle\frac{b}{a} X ~,~
0
\right]
\end{equation}
or equivalently using the ellipticity of the deformation $\beta = (a^2 - b^2)/(a^2 + b^2)$: 
\begin{equation}
\label{eq:libration_base_flow_ellipticity}
\mbf{U}_b^{\ell} =  -\Omega_0 \varepsilon \sin(\Omega_0 f t)
\left[
-(1+\beta) Y ~,~
(1-\beta) X ~,~
0
\right]~.
\end{equation}
The flow (\ref{eq:libration_base_flow_librating_frame})
or (\ref{eq:libration_base_flow_ellipticity})
satisfies the full Navier-Stokes equation in the bulk of the fluid, but is neither a no-slip nor a stress free solution with rigid outer boundary, for which viscous corrections are needed (see for instance the discussion in \cite{sauret_libration-induced_2013}). 
Note that in the absence of elliptical deformation, \ie $\beta = 0$, the flow would be an oscillating solid-body rotation corresponding to the absence of flow in the bulk of the fluid in the mean rotation rate frame of reference.
%

%
The other frame of reference relevant to the study of libration-driven flows is the mean rotation frame, rotating at rate $\Omega_0$.
This frame with axes $(Oxyz)$ is deduced from the preceding one via a rotation around the axis $(Oz)$ of angle $\theta_\ell$ such that:
\begin{equation}
\label{eq:theta}
\theta_\ell (t) = -\Delta \varphi \cos(\Omega_0 f t)   = -\frac{\varepsilon}{f} \cos(\Omega_0 f t)  ~.
\end{equation}
The coordinate change along with velocity composition lead to the following expression for the velocity field at lowest order in $\varepsilon$ \citep{grannan_experimental_2014}:
\begin{equation}
\label{eq:libration_flow_rotating_frame}
\mbf{U}_{b}^{\Omega} ~=~ \Omega_0 \varepsilon \beta \sin(\Omega_0 f t)
\left[
y ~,~
x ~,~
0
\right]~.
\end{equation}
The equation (\ref{eq:libration_flow_rotating_frame})  clearly shows that the libration-driven flow in the mean rotation frame is a periodic, standing strain, whose amplitude is proportional to the ellipticity of the deformation. 
It also leads to defining an input Rossby number $Ro_{i}$ based on the amplitude of the libration flow:
\begin{equation}
Ro_i = \beta \varepsilon \sim \frac{\vert \mbf{U}_b^\Omega \vert}{ \Omega_0 a} ~.
\end{equation}
In the rotating frame of reference, the libration base flow vanishes as $\beta$ goes to zero. 
Lastly, as it will prove useful later, the libration-driven base flow (\ref{eq:libration_flow_rotating_frame}) may also be written in cylindrical coordinates $(r, \phi, z)$ as
\begin{equation}
\label{eq:polar_base_flow}
\mbf{U}_{b}^\Omega = 2 \Omega_0 \varepsilon \beta r  \sin(\Omega_0 f t) \left(\sin 2 \phi \mbf{e}_r + \cos 2 \phi \mbf{e}_\phi \right)
\end{equation}
which shows that it has an $m = 2$ azimuthal wave number, as does the tidal perturbation of the shape of the container.

\subsection{Instabilities of libration-driven flows}

Previous experimental \citep{noir_experimental_2012-2,grannan_experimental_2014}, numerical \citep{cebron_libration_2012,favier_generation_2015}
 and theoretical \citep{cebron_libration-driven_2014} studies have demonstrated that the flow $\mbf{U}_b^{\Omega}$ is unstable. 
The standing oscillating strain is known for driving a parametric subharmonic resonance of inertial modes whose amplitude grows exponentially and which eventually breaks down into turbulence. 
Inertial modes are spontaneous oscillations of rotating fluids caused by the restoring action of the Coriolis force. 
Their existence may be formally deduced from the linearised Euler equation and mass conservation in an incompressible fluid rotating at $\mbf{\Omega}_0$ governing the velocity and pressure fields $\bu$ and $p$:
\begin{equation}
\partial_t \bu  + 2 \mbf{\Omega_0} \times \bu = - \bnabla p ~~~ \mbox{and} ~~~ \bnabla \cdot \bu =  0 
\end{equation}
which can be transformed into the Poincar\'e equation \citep{poincare_sur_1885}:
\begin{equation}
\label{eq:poincare_equation}
\p_{tt} \bnabla^2 \bu + 4 \Omega_0^2 \p_{zz} \bu = 0 ~. 
\end{equation}
This equation has eigenmode solutions $ e^{i \omega t} \mbf{\Phi}_\omega (\mbf{r})$ ($\mbf{r}$ being the position) called ``inertial modes'', whose frequency lies between $-2 \Omega_0$ and $2 \Omega_0$ \citep{greenspan_theory_1968}.
In the parametric subharmonic resonance process, two inertial modes $\mbf{\Phi}_j$ and $\mbf{\Phi}_k$ couple with the base flow $\mbf{U}_b^\Omega$ provided their frequencies match the following condition: 
\begin{equation}
\vert \omega_j - \omega_k\vert = f \Omega_0 ~.
\end{equation}
Their growth rate (normalised by the rotation rate) is then proportional to the input Rossby number $Ro_i = \varepsilon \beta$ \citep{cebron_libration-driven_2014}.   
Note that at finite forcing amplitude, this resonance condition is broadened so that near-resonant modes, that is, modes whose frequencies are such that $\vert \omega_j - \omega_k\vert - f \Omega_0 = O(Ro_i)$ may also be unstable (see for instance \cite{cebron_libration-driven_2014}).

In addition, a significant spatial overlap between the resonant modes and the base flow is required to ensure a fast growth of the modes. 
The base flow has a $m = 2$ azimuthal wave number (see equation (\ref{eq:polar_base_flow})),  which provide an additional resonance condition on the modes' azimuthal structure. 
The azimuthal variations of an inertial mode may be decomposed into a superposition of periodic oscillations with integer wave numbers $m$, \ie a linear combination of functions $\exp (i m \phi)$. 
In a sphere, an inertial mode would be described by a single wave number $m$.
In the present case, since the containers we consider is not invariant under rotation, the radial and azimuthal variations are no longer separable and an infinite superposition of azimuthal wave numbers is require to describe a mode. 
For modes $j$ and $k$ to be resonant, an optimal spatial overlap with the base flow is ensured provided that their most energetic wave numbers, say $m_j$ and $m_k$, satisfy the following relation: 
\begin{equation}
\vert m_j - m_k \vert = 2
\end{equation} 
where $2$ stands for the wavenumber of the libration base flow (see relation (\ref{eq:polar_base_flow})).
In this study, we focus on the libration frequency $f = 4$, for which the growing modes have opposite frequencies such that $\omega_j = - \omega_k = 2 \Omega_0$. 
Both frequencies are at the edge of the inertial wave spectrum.
These modes have a spatial structure made of several horizontal layers with alternating velocities that are perpendicular to the rotation axis (see for instance \cite{favier_generation_2015,grannan_tidally_2017,vidal_inviscid_2017-2,
vidal_diffusionless_2017}).
For these particular modes, the growth rate of the instability is maximal \citep{cebron_libration-driven_2014}, which thus facilitates the exploration of the turbulent saturation of the instability in the low forcing, low dissipation regime. 

\subsection{The saturation of the instability}
\label{sec:theory_saturation}

%
Two possible routes have been foreseen for the saturation of the elliptical instability after previous studies including \cite{barker_non-linear_2013-1,barker_non-linear_2014,
favier_generation_2015,le_reun_inertial_2017-1}. 
As their amplitudes saturate, the inertial modes may transfer energy to daughter inertial modes via a mechanism called triadic resonances \citep{smith_transfer_1999,smith_near_2005,
bordes_experimental_2012-1,lin_experimental_2014}.
The latter are non-linear three modes interactions happening over long time scales for which one inertial mode $i$ gives rise to two other ones $j$ and $k$ provided their frequencies satisfy the following resonance condition: 
\begin{equation}
\label{eq:triadic_resonance_condition}
 \omega_j + \omega_k  =  \omega_i~.
\end{equation} 
Very similarly to the resonance condition for the elliptical instability, at finite (dimensionless) amplitude $A_i$ of the mode $i$, the triadic resonance condition (\ref{eq:triadic_resonance_condition}) is extended to near-resonant modes within a tolerance $O(A_i)$ on the frequencies \citep{smith_near_2005,vanneste_wave_2005}. 
More precisely, the modes $j$ and $k$ are in near-resonant interaction with the mode $i$ when $\omega_j + \omega_k  -  \omega_i = O(A_i)$. 
In the low dissipation and low forcing (or low Rossby number) regime, the successive generation of daughter modes by resonant and near-resonant triadic interactions should give rise to a continuum of inertial waves, with a continuous spectrum of wavelengths and frequencies, a state called ``inertial wave turbulence'' \citep{nazarenko_wave_2011,galtier_weak_2003}
Conversely, the unstable modes could as well lead to the emergence of strong geostrophic vortices, as observed for instance by \cite{barker_non-linear_2013-1} or \cite{le_reun_inertial_2017-1}. 
The vortices correspond to the zero frequency limit of inertial waves, and are quasi-steady flows invariant along the rotation axis. 
As reminded in the introduction, they are ubiquitously observed in rotating turbulent flows, but their emergence out of a set of inertial modes is not fully elucidated. 
It is known for instance that two modes cannot produce a zero frequency mode, or equivalently, that the latter cannot be involved in three modes interaction in the limit of vanishing Rossby and Ekman numbers \citep{greenspan_non-linear_1969}.
Alternative mechanisms have been proposed to explain their presence such as direct non-linear interaction in the boundary layer \citep{tilgner_zonal_2007-1,morize_experimental_2010-1,sauret_tide-driven_2014}, near-resonant interactions \citep{smith_near_2005}
or the geostrophic instability of 
\cite{kerswell_secondary_1999-1}.

Our experimental set-up aims at understanding under which conditions these two possible non-linear saturation states are observed. 
It has been designed to draw closer to low forcing  and low dissipation regimes, which are relevant to planetary interiors.

\section{Experimental methods}
\label{sec:experimental_methods}

\subsection{The experimental set-up}

\begin{figure}
\centering
\includegraphics[width=0.9\linewidth]{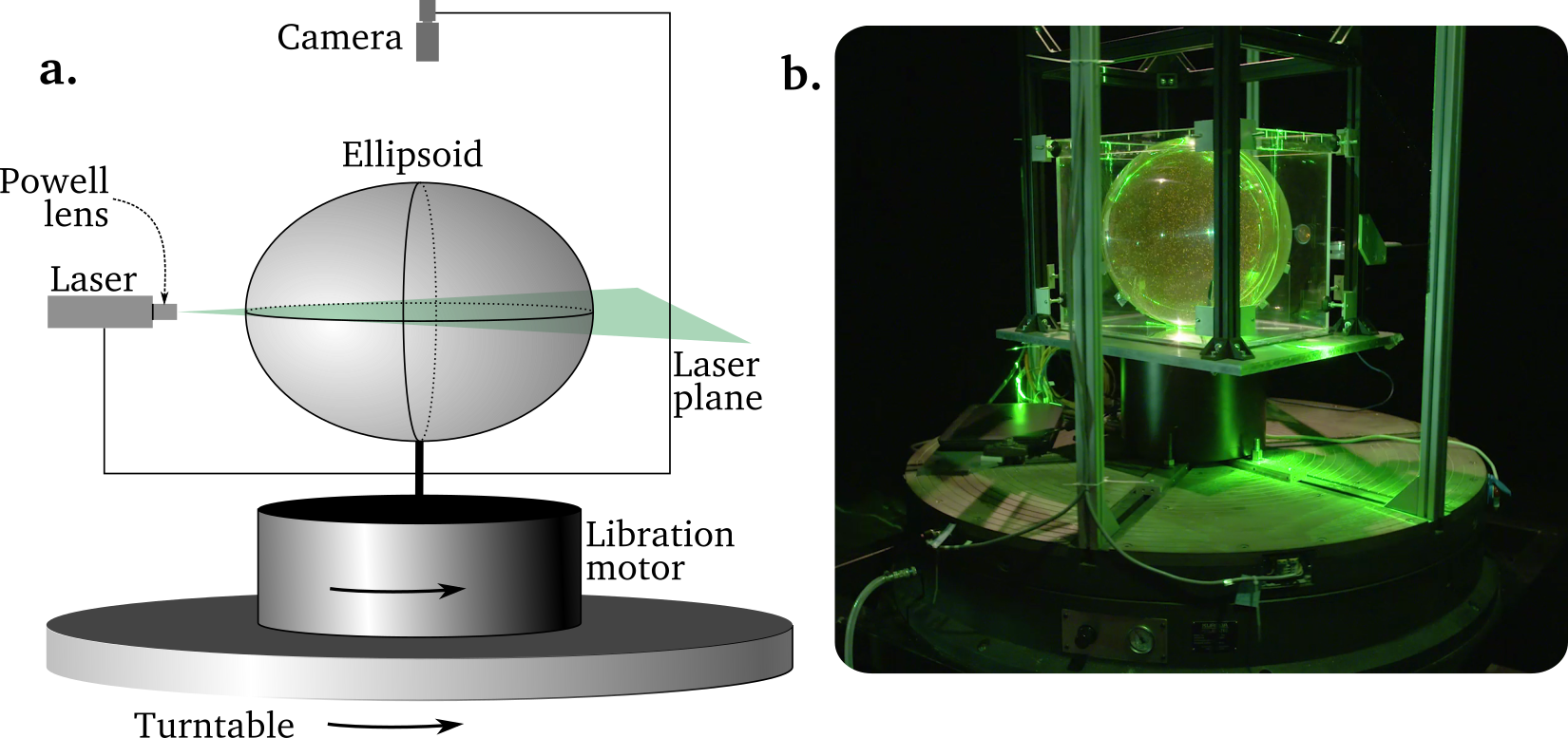}
\caption{
Schematic cartoon (\textbf{a}) and photograph (\textbf{b}) of the experimental setup to implement libration-driven elliptical instability.
A rigid, transparent ellipsoidal container is mounted on a turntable and a secondary oscillating motor, thus producing a rotation rate with harmonic disturbances. The ellipsoid is enclosed in a rectangular box filled with water (not shown on the cartoon for clarity), which limits refraction and ensuing image distortion.
For scale, the vertical extent of the ellipsoid is $2c = 430$ mm.
Note that there was no water in the outer box at the time the picture was taken, and that the Laser sheet was vertical instead of horizontal.
%
}
\label{fig:experimental_setup}
\label{fig:ellipsoid_dimensions}
\end{figure}

\begin{figure}
\centering
\includegraphics[width=0.47\linewidth]{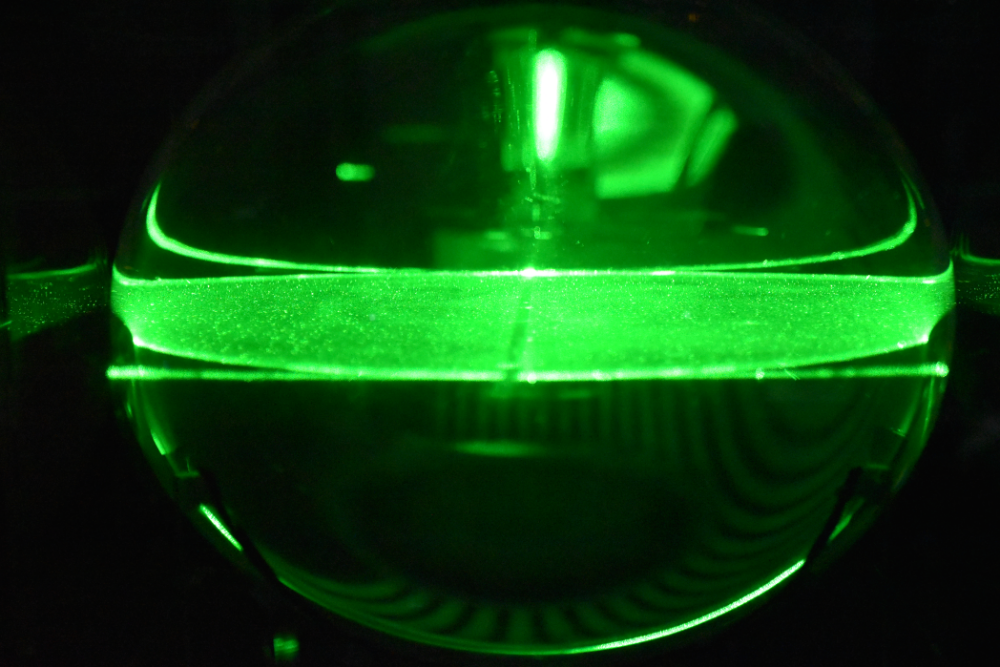}
\hfill
\includegraphics[width=0.47\linewidth]{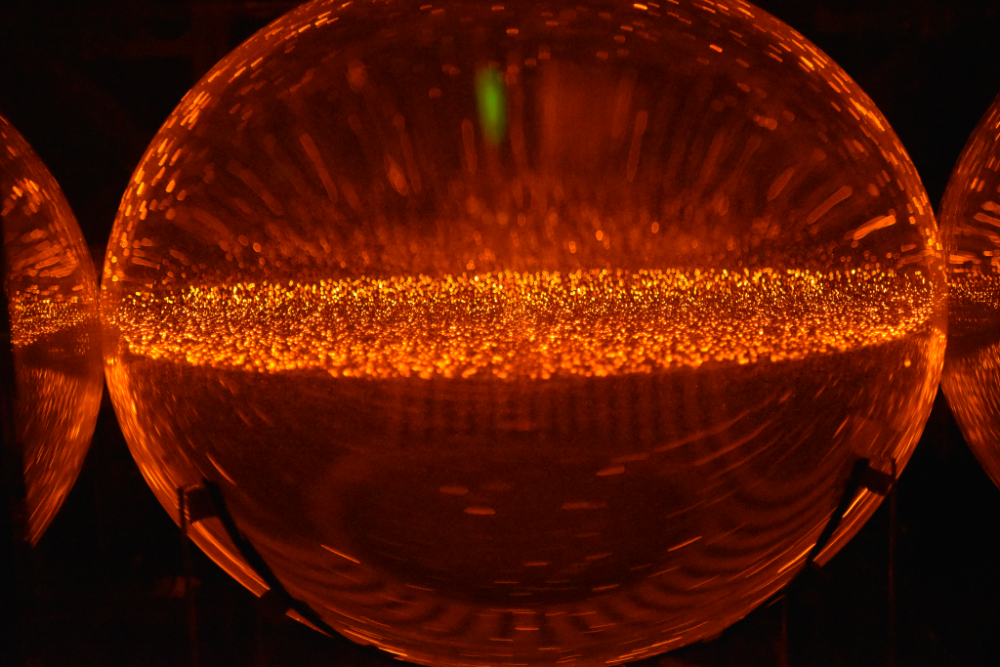}
\caption{Photographs taken without (\textbf{left}) and with (\textbf{right}) an orange filter of the Laser sheet lighting the fluorescent PIV particles. Using PIV particles absorbing the green light and emitting in red allows to filter out the reflections of the sheet on the surface of the ellipsoid, thus improving the overall quality of the images taken by the high-resolution camera. Note that there was no water in the outer box at the time the pictures are taken.}
\label{fig:piv_particles}
\end{figure}

The experiments of libration-driven elliptical instability are carried out in a scaled-up version of the set-up previously implemented by \cite{noir_experimental_2009,noir_experimental_2012-2}, \cite{grannan_experimental_2014} and \cite{grannan_tidally_2017}.
It is depicted in figure  \ref{fig:experimental_setup}. 
%
%
The tri-axial ellipsoid is made of two halves, each one being carved out of a PMMA block down to a sheet of thickness 7 mm with a precision of 0.05 mm, which is about one order of magnitude smaller than the expected boundary layer thickness.
The two sides have been glued together.
%
The lengths of the principal axes of the ellipsoid are $a=254$ mm, $b=178$ mm and $c = 215$ mm, where the axis $c$ is parallel to the rotation axis. 
The ellipticity of the horizontal deformation $\beta$ is therefore  $0.34$.
A 5 mm diameter hole in the ellipsoid allows to fill or empty it with pure water. 
Moreover, the ellipsoid is enclosed in a rectangular box filled with water to reduce optical distortion. 
In the present experiment, the ellipsoid sits on a turntable whose rotation rate ranges from 10 rotations per minute (RPM) to 40 RPM. 
The modulation of the rotation rate $\Omega$ is operated by a $3.1$ kW ring-style, direct-drive, servomotor (\textsc{Yaskawa} SGMCS-2ZN3A11). 
The sinusoidal motion is controlled via a card (Servopack SGD7S) and is discretised on 2000 points per period.
This motor is able to produce oscillations of a system weighting around 100 kg with lateral extent of 30 cm, with a rate ranging from 0 to 180 oscillations per minute. 
The typical crest-to-crest amplitude of the oscillations used in this study ranges from $0^\circ$ to $12^\circ$, although larger amplitudes can easily be reached (the system has been tested up to $50^\circ$ at $0.5$ Hz).
Relative variations of the rotation rate of the primary turntable due to the oscillations of the servo-motor do not exceed $0.4$~\%. 
The range of turntable rotation rates and libration amplitudes are indicated in table \ref{tab:runs_statistical}.
To measure the velocity of the flow, we use Particle Image Velocimetry (PIV). 
A green Laser beam is transformed into a 2 mm-thick homogeneous sheet with a Powell lens; the sheet cuts the ellipsoid through the equatorial plane (see figure \ref{fig:experimental_setup}) and lights PIV particles with a fluorescent coating (Cospheric UVPMS-BO-1.00, 53-63$\mu$m). 
The fluorescent coating is such that it absorbs the green light to emit in orange; filtering out the green light allows to capture only the particles and to get rid of parasitic reflections  of the sheet on the ellipsoid (see figure \ref{fig:piv_particles}).
High resolution images ($2560\times 1600$ pixels) of the Laser plane are taken with a camera (\textsc{Dantec} SpeedSense 341)  placed above the ellipsoid, the CCD sensor being at a distance of $38.7$ cm from the equatorial plane.
It is mounted with a 28 mm Zeiss lens. 
The camera is attached to a structure bound to the secondary motor, so that a fixed-shape ellipsoid is seen through the camera. 
The framerate used in our experiments ranges from 20 frames per second (fps) for the lowest rotation rate, to 50 fps for the highest. 
Image acquisition is controlled by \textsc{Dantec}'s software DynamicStudio. 
Note that the duration of a recording is limited by the available memory in the camera (32 Gbytes). 
The data transfer from the camera to the computer must be completed before a new recording is started, which hence prevents long and continuous acquisitions. 
The DynamicStudio software is used to process the images into velocity fields, via an adaptive PIV algorithm.
The PIV algorithm is performed on $78 \times 50$ boxes of $64 \times 64$ pixels size. 
Note that the number of PIV particles in the ellipsoid is such that there is always about $3$ to $5$ particles inside these boxes.
Lastly, both the camera and the servomotor are controlled by a computer lying on the turntable, which is monitored remotely via WiFi.

The calibration of the PIV field of view could not be achieved using a classical grid pattern, as the interior of the ellipsoid can be accessed only via a 5 mm hole. 
Instead, we have developed a non-invasive method based on the creation of a light pattern and on numerical ray path construction.
This special method which provides a satisfying calibration is detailed in  Appendix \ref{appA}.

A summary of the experimental parameters is given in table \ref{tab:runs_statistical}.
We present in particular the values of the libration amplitude, the input Rossby number and the Ekman number which compares the effects of viscosity and Coriolis force. 
We emphasise the fact that  our set-up allows  to reach values of the Ekman number that are about an order of magnitude below those of \cite{grannan_experimental_2014}.
This is made possible by the increase in the size of the ellipsoidal container, and the use of the most powerful servo-motor of its category.

\begin{table}
  \begin{center}
\def~{\hphantom{0}}
  \begin{tabular}{cccccc}
    Rotation rate  & Ekman number $E$ & Libration angle range &  $\varepsilon$ range   &   
    $Ro_i \times 10^{2}$ range   \\[3pt]
      10 RPM & $1.5 \times 10^{-5}$ & $2.64^\circ$--$4.74^\circ$   & $0.185$--$0.331$ & $6.31$--$11.2$   \\
      20 RPM &  $7.4 \times 10^{-6}$ & $1.68^\circ$--$4.22^\circ$   & $0.118$--$0.294$ & $4.02$--$10.0$   \\
      30 RPM &  $5.0 \times 10^{-6}$ & $1.07^\circ$--$4.23^\circ$   & $0.0747$--$0.296$ & $2.54$--$10.1$  \\
 	  40 RPM &  $3.7 \times 10^{-6}$ & $1.41^\circ$--$1.74^\circ$   & $0.0986$--$0.122$ & $3.36$--$4.16$  \\

  \end{tabular}
  \caption{
Summary of the different input parameters used in the experiment and the corresponding dimensionless parameters. The rotation rate is in rotations per minute (RPM). The Ekman number  is defined as $E = \nu /(a^2 \Omega_0) $, where $\nu$ is the viscosity of water, considered to be $1.0 \times 10^{-6}$ m$^2$.s$^{-1}$ at room temperature; it compares the effects of viscosity to those of the Coriolis force.
Note that the typical relative uncertainty on the value of the libration angle is about $1.5$ \%.  
}
  \label{tab:runs_statistical}
  \end{center}
\end{table}

\subsection{The experimental procedure}
\label{sec:experimental_procedure}

To study the non-linear saturation of the elliptical instability, we proceed as follows.
We first turn on the turntable and wait for the fluid inside the ellipsoid to reach solid-body rotation.
We then turn on the secondary motor which imposes oscillations at a dimensionless rate $f=4$ with a desired amplitude $\Delta \varphi$. 
Between two experiments carried out at the same rotation rate, the secondary motor is turned off, and we wait for the remaining flow inside the ellipsoid to dissipate before starting a new run.

\section{Base flow and frames of reference}

\subsection{Measuring the libration flow}

In order to characterise each experiment and to quantify the flows excited by the elliptical instability, the amplitude and the phase of the libration base flow must be precisely known.
These two quantities are determined from the experimental data in the early phase of each experiment when libration is imposed but no instability has significantly grown yet: the experimental base flow is fitted to the analytical formula (\ref{eq:libration_base_flow_librating_frame}) in order to determine the amplitude and phase.
We detail this process in the following and also illustrate the agreement between the theoretical expression and the experimental measurement of the libration base flow. 

%
%

\begin{figure}
\includegraphics[width=\linewidth]{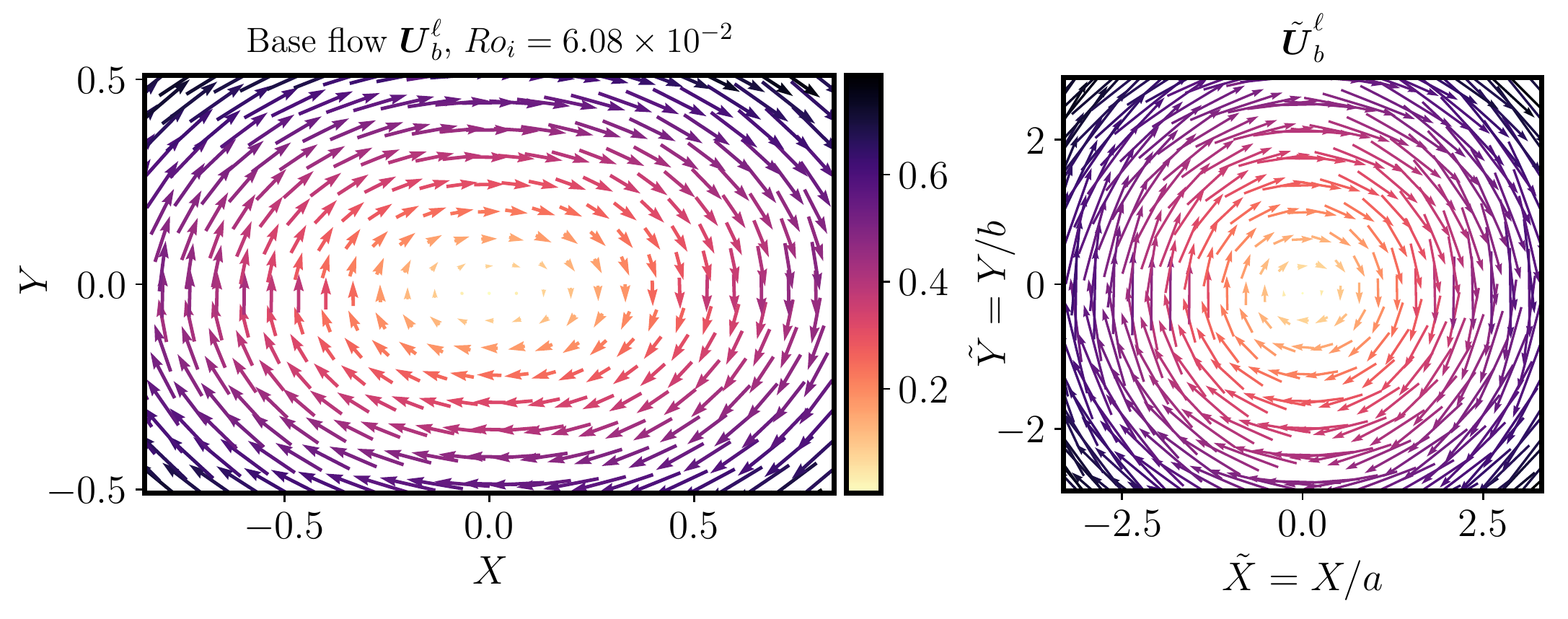}
\caption{Experimental measurement of the libration base flow $\mbf{U}_b^\ell$ (\textbf{left}) and the transformed base flow $ \tilde{\mbf{U}}_b^\ell$ (\textbf{right}) for an input Rossby number $Ro_i = 6.08 \times 10^{-2}$. Note that $X$ and $Y$ are normalised by $a$, and that the velocity is scaled by the typical libration velocity $Ro_i a \Omega_0$. }
\label{fig:base_flow}
\end{figure}

%
From measurements performed in the librating frame, we
first define a transformed base flow $\tilde{\mbf{U}}_b^\ell = (\tilde{U},\tilde{V})$ and a transformed position $\tilde{\mbf{X}} = (\tilde{X},\tilde{Y})$ such that:
\begin{equation}
\label{eq:coordinate_transformation}
\tilde{U} = U/a ~,~ \tilde{V} = V/b ~,~ \tilde{X} = X/a ~,~ \tilde{Y} = Y/b 
\end{equation} 
where $U$ and $V$ are respectively the $X$ and $Y$ components of the measured libration base flow in the libration frame of reference.
%
%
Applying this transformation to the theoretical base flow (\ref{eq:libration_base_flow_librating_frame}) yields:
\begin{equation}
\label{eq:transformed_base_flow}
\tilde{\mbf{U}}_b^\ell = \frac{-2ab}{a^2 + b^2} \Omega_0 \varepsilon \sin(\Omega_0 f t)
\left[
- \tilde{Y} ~,~
\tilde{X} ~,~
0
\right] =  \frac{-2ab}{a^2 + b^2} \Omega_0 \varepsilon \sin(\Omega_0 f  t) \tilde{r} \tilde{\mbf{e}}_{\phi} = \tilde{\Omega} (t)\tilde{r} \tilde{\mbf{e}}_{\phi}
\end{equation}
which is an oscillating solid-body rotation with effective rotation rate $\tilde{\Omega}$, and where we have introduced $\tilde{r} =\sqrt{\tilde{X}^2 + \tilde{Y}^2} $ and $\tilde{\mbf{e}}_{\phi}$ an orthoradial vector in the transformed coordinates. 
Experimental PIV fields of the base flow $\bU_b^{\ell}$ 
is shown in figure \ref{fig:base_flow} along with the flow transformed according to (\ref{eq:coordinate_transformation}).
They are in good agreement with the theoretical expressions (\ref{eq:libration_base_flow_librating_frame}) and (\ref{eq:transformed_base_flow}).

%

\begin{figure}
\includegraphics[width=\linewidth]{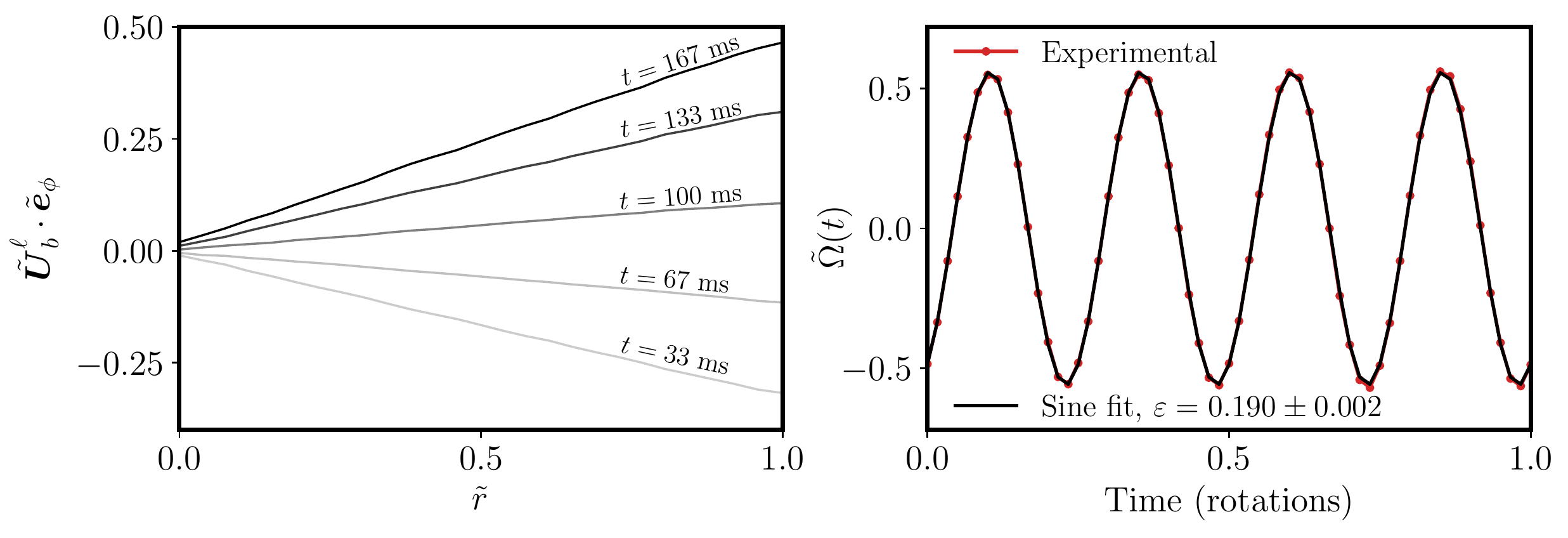}
\caption{\textbf{Left:} experimental measurement of $\tilde{\mbf{U}}_b^\ell \cdot \tilde{\mbf{e}}_{\phi}$ over half a period of libration. 
Each curve is labelled by the time $t$ at which the quantity is determined and is given in milliseconds.
\textbf{Right:} experimental measurement of the effective rotation rate $\tilde{\Omega}$ defined in   equation (\ref{eq:transformed_base_flow}) and best sinusoidal fit yielding $\varepsilon = 0.190 \pm 0.002$ in this case. The fitting parameters are the phase and the amplitude of the sine. 
The agreement between the fit and the experimental data is within 1.5 \% . 
}
\label{fig:base_flow_properties}
\end{figure}

%
%
To determine the axis of rotation, the field $ (\tilde{\mbf{U}}_b^\ell)^2$ is fitted to an axisymmetric parabola with adjustable central position; the location of the center is then averaged over the whole set of velocity fields.
The same transformed field is used to determine $\varepsilon$ and the phase of the libration forcing: the position is sorted in 30 rings centred on the axis of rotation, on which the orthoradial transformed velocity is averaged.
The result of this process is shown over half a libration period in the left panel of figure \ref{fig:base_flow_properties}.
Fitting the averaged orthoradial velocity with a line gives the effective rotation rate $ \tilde{\Omega}(t)$ which is represented in the right panel of figure \ref{fig:base_flow_properties}. 
The time series of $\tilde{\Omega}$ is fitted to a sinusoidal function to measure the amplitude of libration $\varepsilon$ and the phase. 
The agreement between the experimental data and the sinusoidal function is within 1.5~\% relative error.

\subsection{Transformation from the libration to the rotating frame}
\begin{figure}
\centering
\includegraphics[width=\linewidth]{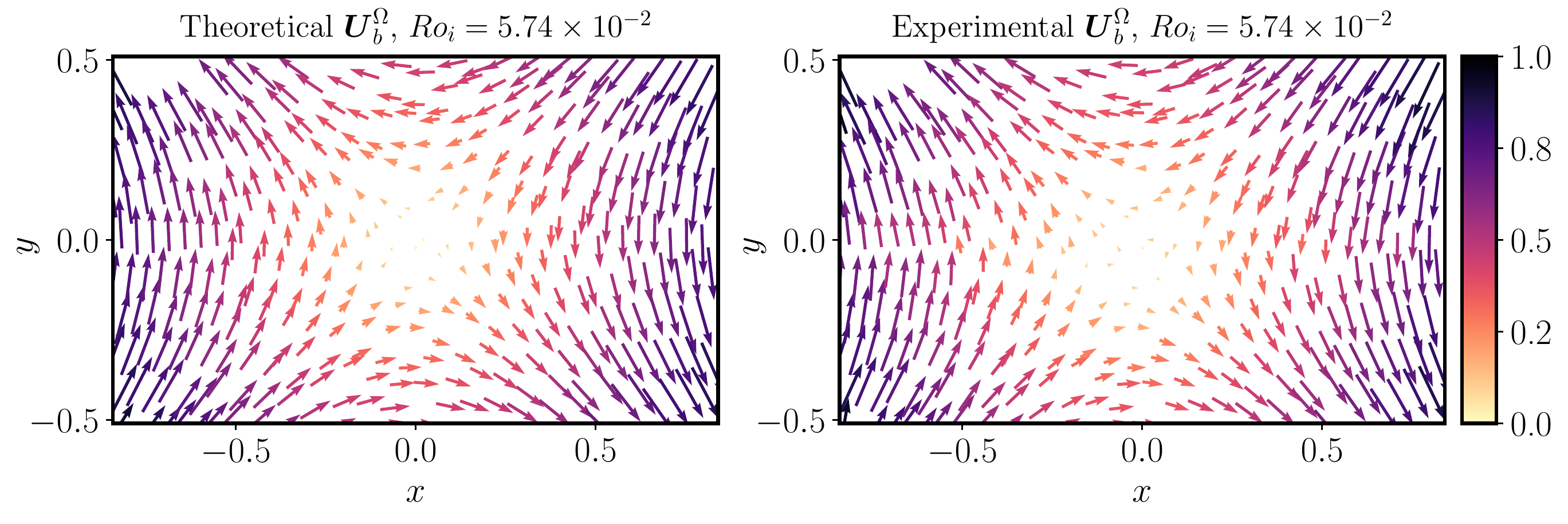}
\caption{Snapshot of the theoretical (\textbf{left}) and experimental (\textbf{right}) libration base flow $\bU_b^\Omega$ transformed into the rotating frame, for an input Rossby number $Ro_i = 5.74 \times 10^{-2}$. The theoretical base flow is given by formula (\ref{eq:libration_flow_rotating_frame}). In this frame, the libration base flow is a standing strain field. The velocity is normalised by a typical libration velocity $Ro_i a \Omega_0$ and the distances are normalised by $a$. 
}
\label{fig:rotating_frame_base_flow}
\end{figure}

Although the camera is in the librating frame of reference for experimental convenience, the adequate frame to study the dynamics of the non-linear saturation of the elliptical instability is the rotating frame in which inertial waves and the geostrophic modes are well defined. 
The transformation of the PIV fields from one frame to another is performed in the post-processing phase. 
The position in the rotating frame is deduced from the position in the libration frame by a rotation of angle $\theta_\ell$ as defined in (\ref{eq:theta}).
The velocity at this rotating position is computed from interpolation of the PIV field with 3rd order two-dimensional splines using the \texttt{RectBivariateSpline} function of the \textsc{Python} library \texttt{SciPy} \citep{jones_scipy_2001-1}.
This transformation also includes a velocity composition: a solid body rotation associated to the rotation of varying angle $\theta_{\ell}$ is removed from the measured velocity. 
A snapshot of the base flow transformed into the rotating frame is shown in figure \ref{fig:rotating_frame_base_flow}.
It is compared to the theoretical base flow at the same time and the overall agreement between the two fields is satisfactory. 
Discrepancies may be noticed around the line $y= 0$ that are due to the line where the two parts of the ellipsoid are glued together and where the optical distortion is important. 
Other discrepancies may be noticed in the centre where the particle displacement are small and the direction of their motion is therefore difficult to determine from the PIV algorithm.
Nevertheless, the relative error between the theoretical and the experimental base flow in the rotating frame computed with $L^2$ norm is as low as $3$ \%.   

%


\section{The non-linear saturation at low forcing amplitude: waves in interaction}
\label{sec:low_forcing_amplitude}

In the following, we focus on the non-linear saturation of the elliptical instability driven by libration at low forcing amplitudes.
For a fixed rotation rate, or Ekman number $E$, and libration frequency, the procedure described in paragraph \ref{sec:experimental_procedure} is repeated with increasing libration amplitude or input Rossby number $Ro_i$.
We start with amplitudes that are below the threshold of the elliptical instability, which is approximately found at $Ro_i \simeq 10 \sqrt{E}$ (a value rather similar to \cite{favier_generation_2015}), and then explore the non-linear regimes.
We first present time series of the saturation flow which indicate that despite the libration amplitudes being small, we observe a chaotic, if not turbulent, state.
We then explore the spectral content of the flow and identify inertial waves that are in triadic resonant interaction with one another. 
What we mean by ``low forcing amplitude'' is let unspecified in this section, the boundaries of the regime we detail here will be explored afterwards. 


\subsection{Kinetic energy time series}
\label{sec:kinetic_energy}

To describe in general the saturation of the libration-driven elliptical instability, we track the evolution of the kinetic energy over time. 
In particular, we focus on the kinetic energy of the fluctuations around the libration base flow. 
These fluctuations are measured directly in the libration frame, \ie there is no need to proceed to a change of frame of reference to carry out this measurement. 
Let us call $\bu$ the fluctuations, such that the total flow in the libration frame of reference writes:
\begin{equation}
\label{eq:total_flow_libration}
\bU (\bX) = \BFL (\bX) + \bu(\bX)~.
\end{equation}
The flow in the rotating frame is deduced from (\ref{eq:total_flow_libration}) by a coordinate rotation $\bX \mapsto \bx$ and a velocity composition with the solid body rotation
$$\bU_{\mathrm{rot}} = \Omega_0 \varepsilon \sin(\Omega_0 f t) \left[-y, x, 0 \right] $$ 
such that: 
\begin{equation}
\bU (\bx) = \BFL(\bx) - \bU_{\mathrm{rot}}(\bx) + \bu (\bx) = \BF(\bx) + \bu (\bx)~, 
\end{equation} 
where $\BF(\bx)$ is the base flow measured in the mean rotation rate frame of reference.
It is therefore equivalent to measure the kinetic energy of the fluctuations in these two frames. 
We therefore define a fluctuation kinetic energy $\mathcal{E}$ and a base flow kinetic energy $\mathcal{E}_b$: 
\begin{equation}
\label{eq:Eb_definition}
\mathcal{E} =\frac{1}{2}\left\langle \bu^2 \right\rangle~~~ \mbox{and} ~~~ \mathcal{E}_b =  \frac{f \Omega_0}{2 \pi} \int_{0}^{\frac{2\pi}{f \Omega_0}} \left\langle  \frac{1}{2}   (\BF)^2 \right\rangle \, \mathrm{d}t
\end{equation}
where the operation $\left\langle \cdot \right\rangle$ denotes a summation over the all the points of the velocity field obtained from the PIV measurements, located in the equatorial plane of the ellipsoid.
%

%
\begin{figure}
\centering
\includegraphics[width=\linewidth]{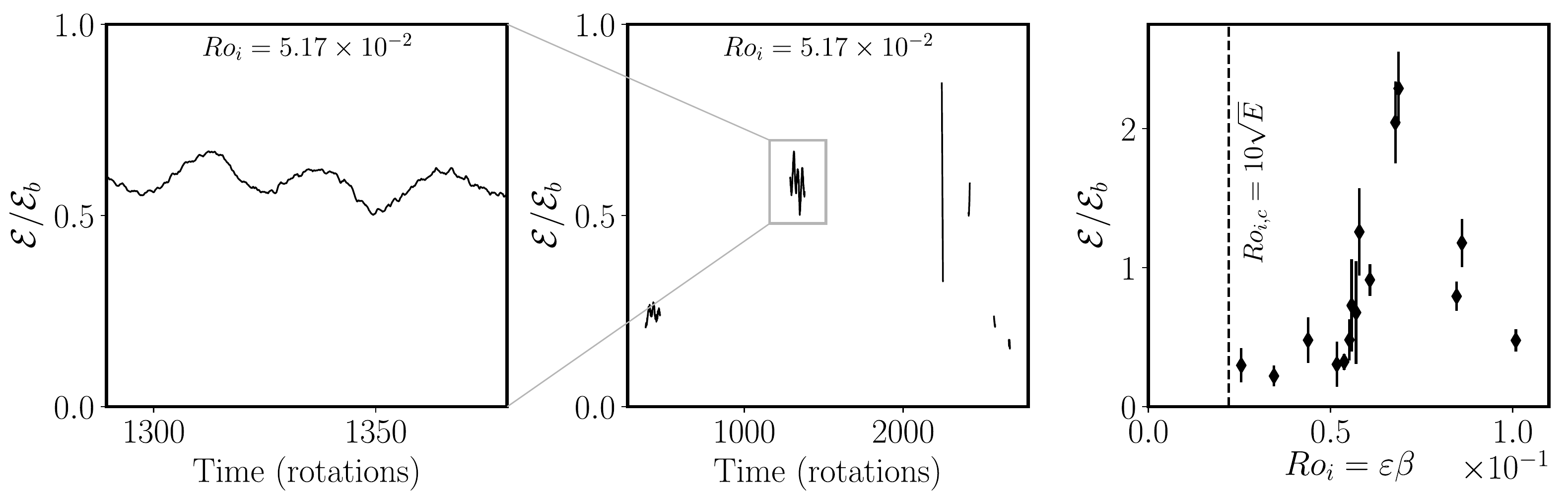}
\caption{\textbf{Left}: kinetic energy variation over a hundred of rotation times at an Ekman number $E = 5.0 \times 10^{-6}$, $f = 4$ and $Ro_i = 5.17 \pm 0.07 \times 10^{-2}$. 
\textbf{Centre}: several successive time series of the kinetic energy for the same experimental run as in the left panel.
This panel highlights the long-time variability of the saturation flow.
The minimum separation between two successive recordings of the velocity field is set by the time needed to transfer the data from the camera to the computer. 
This data set comprises long (5476 images) and short (500 images) recordings at a framerate of $30$ fps. %
Note that these time series include the one shown in the left panel.
For both panels, the time series are low-pass filtered by a sliding average over one rotation period. 
\textbf{Right}: mean value of the kinetic energy in the saturation phase of the instability at $30$ RPM and $f = 4$ for all the experimented values of the input Rossby number $Ro_i$. 
The error bar is given by the standard deviation of kinetic energy considering all successive recordings for each experiment (see for instance the central panel). 
The vertical dashed line materialises the approximate viscous threshold of the instability according to \cite{le_bars_tidal_2010} and \cite{favier_generation_2015}, which is $Ro_{i,c} \sim 10 \sqrt{E}$, and below which the flow is experimentally observed to be stable.
 }
\label{fig:kinetic_energy}
\end{figure}
The result of computing the ratio $\mathcal{E}/ \mathcal{E}_b$ is shown in figure \ref{fig:kinetic_energy} where we represent time series of this quantity and its mean value for several experiments at an Ekman number $E = 5.0 \times 10^{-6}$ (see table \ref{tab:runs_statistical}).
The quantity $\mathcal{E}$ undergoes fluctuations over time, be it at time scales as short as the rotation period or at very long times scales, similar to the slow non-linear time scale $2 \pi (Ro_i \Omega)^{-1}$ ranging from $60$ to $200$ rotation periods.
In addition, the ratio $\mathcal{E}/\mathcal{E}_b$ is of order 1, which indicates that the amplitude of the fluctuations is order $Ro_i$. 
This result is expected from balancing the non-linear term $\vert \bu \cdot \bnabla \bu \vert \sim u^2 / \lambda $ (where $\lambda$ is the resonant mode wavelength) with forcing terms $\vert \BF \cdot \bnabla \bu + \bu \cdot \bnabla \BF\vert \sim Ro_i u /\lambda  $ leading to a saturation velocity perturbation $u \sim Ro_i$. 
A similar saturation amplitude was already observed in previous experiments and simulations \citep{barker_non-linear_2013-1,grannan_tidally_2017}.
These two features of the kinetic energy attest that the saturation phase is dominated by non-linear  transfers, and is potentially turbulent. 
Variations around the simple scaling $ \mathcal{E}/\mathcal{E}_b =  O(1) $ are noticeable in figure \ref{fig:kinetic_energy} but remain difficult to explain since our set-up only allows to measure horizontal motions in a plane. 
In particular, the vertical motions along the rotation axis and their dependence with $Ro_i$ are not accessible.
In this section, we are interested in the saturation flow for forcing amplitudes that are above the threshold of the instability but which remain relatively small (typically $Ro_i < 5.58 \times 10^{-2}$ at $E = 5.0 \times 10^{-6}$). 
The right panel of figure \ref{fig:kinetic_energy} proves that even close to the threshold the kinetic energy of the perturbation flow $\bu$ reaches significant values ($\mathcal{E}/\mathcal{E}_b > 0.3$) and undergoes temporal variability. 

\subsection{Spectral content of the flow: inertial waves in resonant interaction}
\label{sec:wave_dominated_spectral_content}

To refine the analysis of the non-linear saturation flow we explore its spectral content in the temporal domain. 
It is a natural analysis to perform because many features of the flow have a specific frequency signature: the libration forcing appears at $\omega = f = 4$, and inertial waves are restrained to the domain $\left[ - 2, 2 \right]$ (time being normalised by the rotation time scale $\Omega_0^{-1}$).
To determine the spectral content of the flow, we randomly select a set of 300 locations in the rotating frame of reference where the two components of the velocity are recorded. 
Prior to performing the temporal Fourier transform, the local velocity time series are multiplied by the Hann windowing function $\mathcal{H}$ defined as follows $$ \mathcal{H}(t) = \frac{1}{2} - \frac{1}{2} \cos \left( \frac{2 \pi}{T} t\right), $$
where we have assumed that the local velocity time series span over a time interval $\left[ 0, T\right]$.
This windowing avoids the emergence of artificial secondary maxima around peaked frequencies, due to the non-periodicity of the experimental signals. 
Lastly, we perform an ensemble average over the set of 300 power spectra, yielding a quantity denoted as $\hat{E}(\omega)$. 
These power spectra are scaled by a typical local energy of the base flow $e_b$ that is defined as the total base flow kinetic energy $\mathcal{E}_b$ divided by the number of PIV boxes composing the field of view. 
%

%
\begin{figure}
\centering
\includegraphics[width=0.55\linewidth]{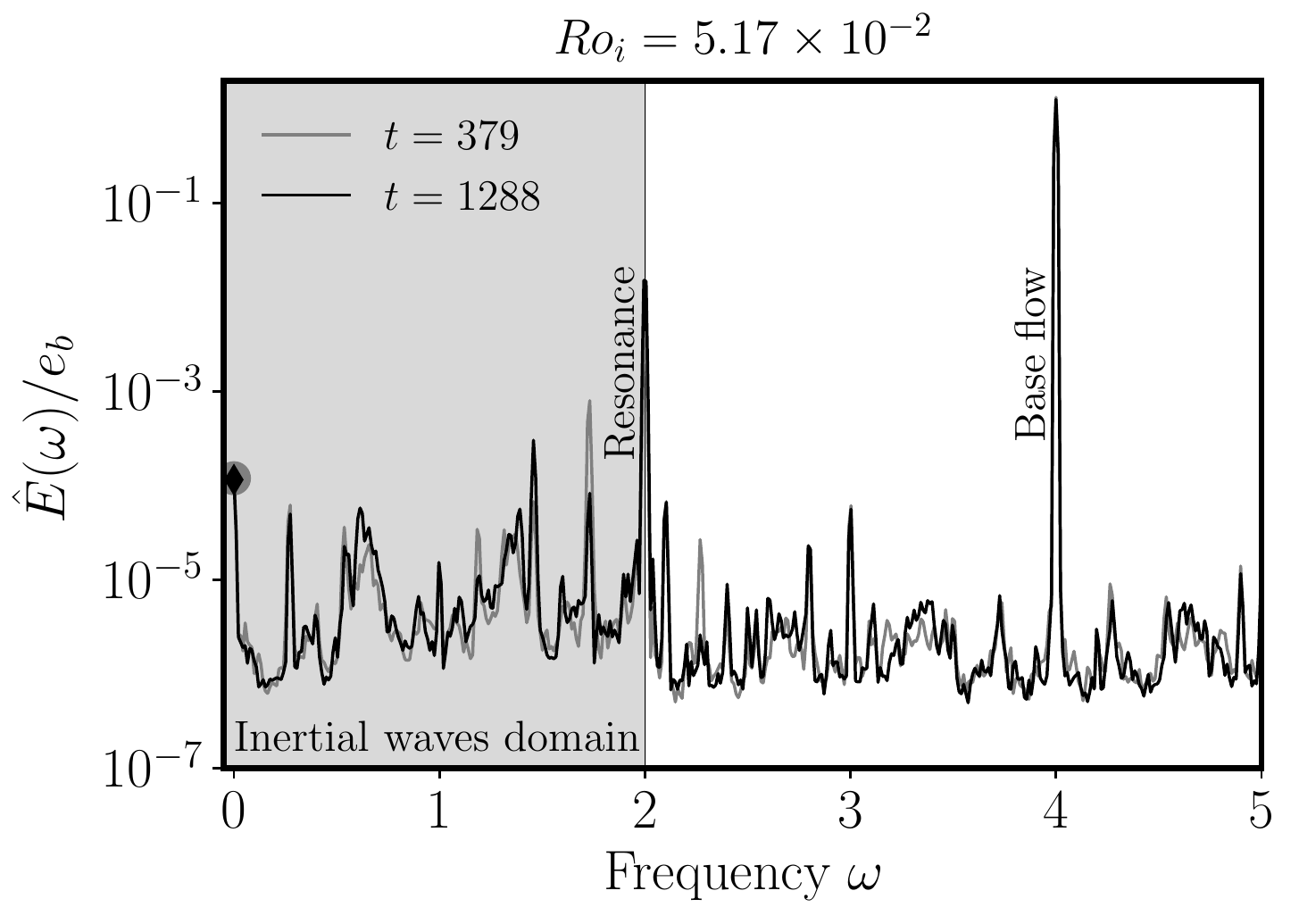}
\caption{Ensemble average of temporal spectra at a forcing amplitude $Ro_i = 5.17 \times 10^{-2}$, a libration frequency $ f =4$ and an Ekman number $E= 5.0 \times 10^{-6}$. 
The frequency $\omega$ is given in rotation units, so that $\omega = 1$ corresponds to a frequency of $\Omega_0$.
Two different power spectra are displayed, they correspond to two different recordings around a time $t$ indicated in rotation periods in order to enhance robust features. 
Among the clear peaks are the base flow at $\omega = f =4$ peaking at $\hat{\mathcal{E}}(f)/e_b \sim 1$, and the resonant mode at $\omega = f/2 = 2$. 
The shaded area denotes the domain of existence of inertial waves.
The two symbols at $\omega = 0$ highlight the value taken by both spectra at this frequency, their color being the same as the corresponding spectra. 
The typical frequency resolution is about $1.1 \times 10^{-2}$. 
}
\label{fig:O30RPM_low_forcing}
\end{figure}

\begin{figure}
\includegraphics[width=\linewidth]{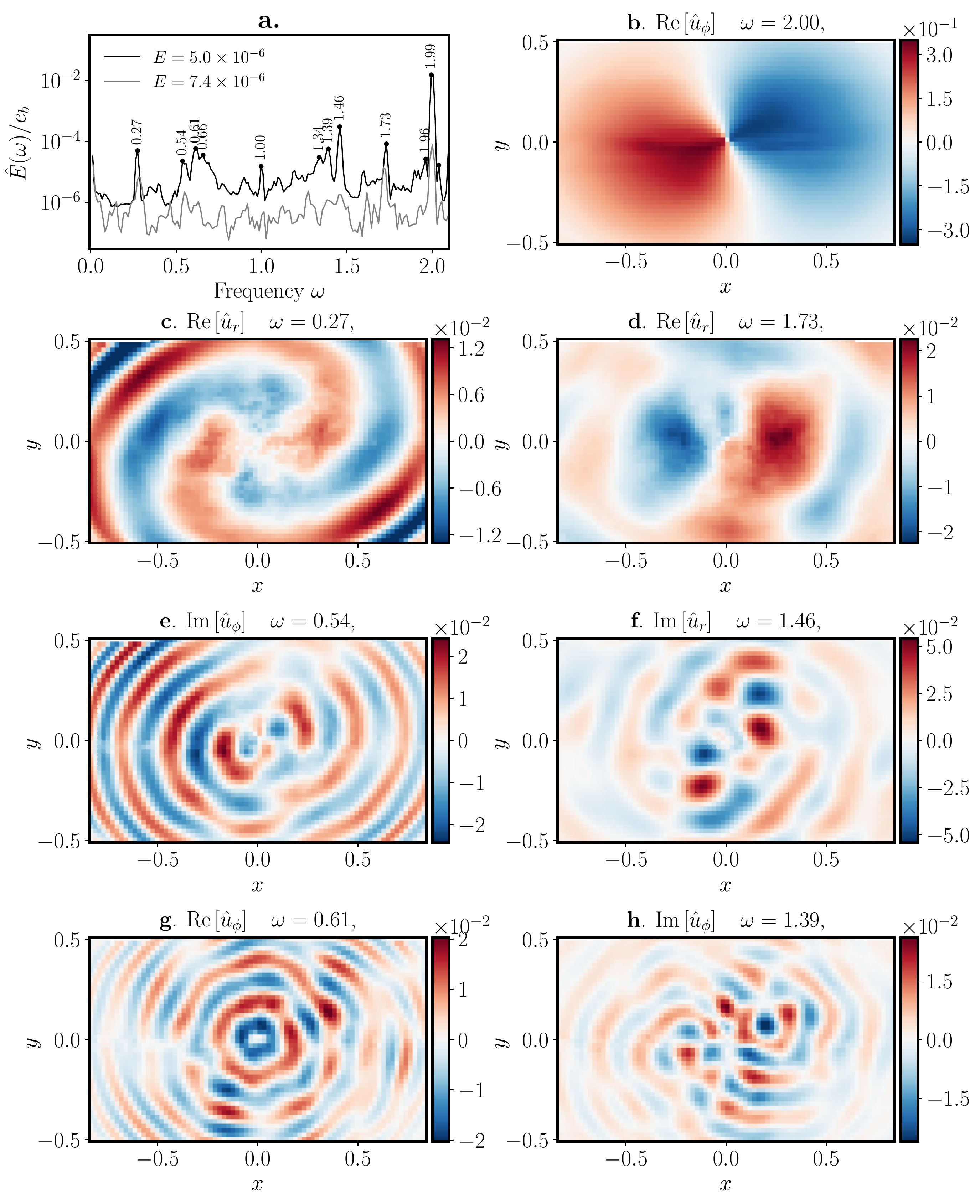}
\caption{ \textbf{a.} Power spectra with labelled peaks. The spectrum at $E= 5.0 \times 10^{-6}$ is the same as in figure \ref{fig:O30RPM_low_forcing} ($Ro_i = 5.17 \times 10^{-2}$). 
The $E= 7.4 \times 10^{-6}$ spectrum is determined from an experiment at input Rossby number $Ro_i = 5.90 \times 10^{-2}$, it has been vertically shifted to facilitate the comparison.
Several pairs of triadic resonance relation can be noticed, such as $0.27 + 1.73$, $0.54 + 1.46$, $0.61+ 1.39$ and $0.66 + 1.34$. 
\textbf{b:}  azimuthal component of the resonant wave. \textbf{c to g:} pairs of structures $\hat{\bu}(\mbf{r}; \omega)$ satisfying the triadic resonance condition on the frequency (relation (\ref{eq:triadic_resonance_condition})).
For each frequency, including $\omega = 2$, we show the component which has the largest amplitude among the imaginary and real parts of the radial and azimuthal velocity.
Lastly, in all velocity maps, distance is normalised by $a$ and velocity by $Ro_i a \Omega_0$.
}
\label{fig:triadic_resonance}
\end{figure}

A typical result of this process for an Ekman number $E = 5.0 \times 10^{-6}$, a libration frequency $f = 4$ and an input Rossby number $Ro_i = 5.17 \times 10^{-2}$  is shown in figure \ref{fig:O30RPM_low_forcing}.
As anticipated, the peaks associated with the libration flow ($\omega = f = 4$) and the resonant modes ($\omega = f/2 = 2$) appear clearly in the power spectra.
In addition, superposing power spectra for the same experiment but at different times reveals the overall reproducibility of this measurement. 
The peaks, in particular in the inertial modes range, and the low background noise, indicate that the saturation at low forcing is chaotic or weakly turbulent, and includes persistent inertial modes. 
Furthermore, we notice in figure \ref{fig:triadic_resonance}a that their position is not random but persists at different Ekman numbers provided the forcing remains small.
Besides, the peaks can be paired together so that their frequencies $\omega_1$ and $\omega_2$ satisfy:
\begin{equation}
\omega_1 + \omega_2 = 2~.
\end{equation}
This suggests that the paired structures are inertial waves in triadic resonant interactions with the resonant mode, as explained in subsection \ref{sec:theory_saturation} (see in particular relation (\ref{eq:triadic_resonance_condition})).

\begin{figure}
\includegraphics[width=\linewidth]{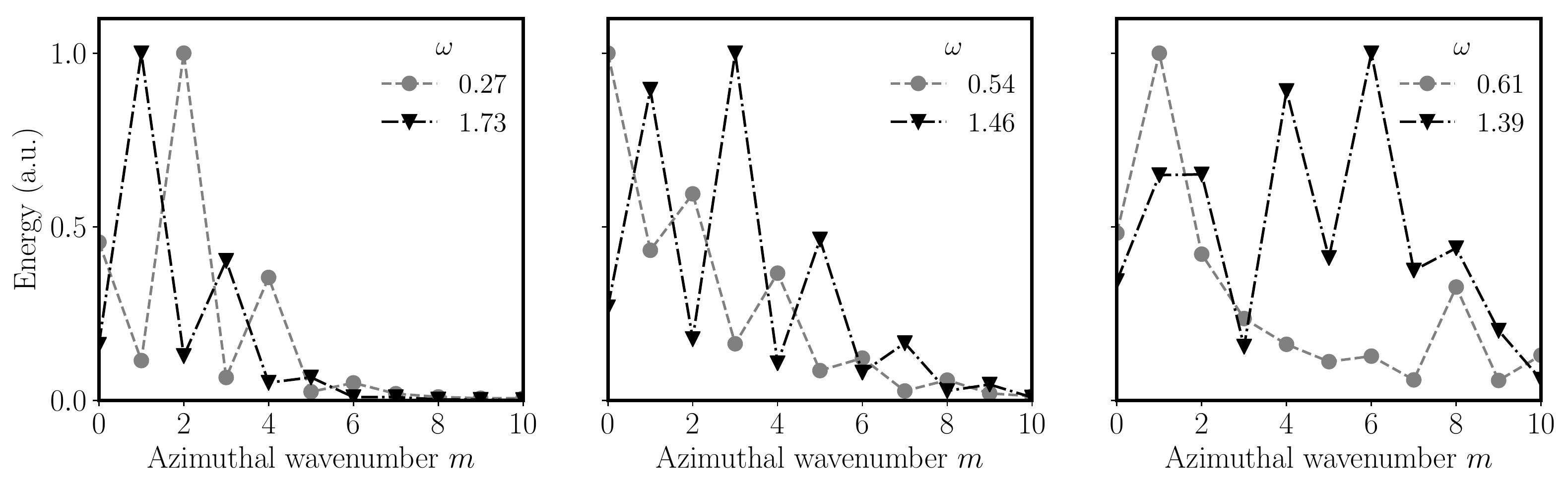}
\caption{Computation of the azimuthal wavenumber decomposition (see formulae (\ref{eq:azimuthal_dec}) and (\ref{eq:azimuthal_mode_dec_energy})) of the paired structures shown in figure \ref{fig:triadic_resonance}. The energy contained in the wave numbers is shown in arbitrary units. Note that these maps are symmetric respective to $m = 0$, so that the energy contained in the $m$ and $-m$ modes is the same. For these structures to significantly interact with the resonant mode, the gap between the displayed principal wavenumbers numbers must be $\pm 1$ (see relation (\ref{eq:resoance_wave_number})), which is indeed observed for the first two pairs, but less obvious for the last one. Note that in the computation of the energy, the radial integration has been restricted to radii below $0.5 a$, which corresponds to the radius of the largest circle fitting inside the PIV fields.}
\label{fig:azimuthal_wave_number}
\end{figure}

To further support our claim that the temporal power spectra present evidence for triadic resonant interactions, we extract from the PIV fields the structure at the peaked frequency. 
The structure oscillating at frequency $\omega$, \ie the Fourier component of the flow at this frequency $\hat{\bu}(\omega)$, is extracted using the following operation:
\begin{equation}
\label{eq:structure_Fourier_component}
\hat{\bu}(\mbf{r};\omega) ~=~ \int_{t_0}^{t_1} \bU (\mbf{r} ,t) e^{i \omega t} ~ \mathrm{d} t~, 
\end{equation}
where $t_0$ and $t_1$ are the start and end times of a PIV recording and $\bU$ is the total velocity field measured in the rotating frame.  
%
%
The result of such a process is shown for several pairs of frequencies  in figure \ref{fig:triadic_resonance}. 
The spatial structures extracted at the peaked frequencies show a spatially periodic organisation which is typical of inertial modes. 
Comparison with theoretical structures remains nevertheless complicated because the viscous inertial modes of a tri-axial ellipsoid are not known theoretically and require complicated numerical computation \citep{vidal_inviscid_2017-2,vidal_diffusionless_2017}. 
In addition, the ability of the paired modes to interact with the resonant one may be assessed by examining their principal azimuthal wave numbers. 
The latter is determined for a mode $\hat{\bu}(\mbf{r};\omega)$ by transformation into polar coordinates $(r, \phi)$, and Fourier transform along $\phi$ to obtain the following decomposition of the mode:
\begin{equation}
\label{eq:azimuthal_dec}
\hat{\bu}(r,\phi;\omega) = \sum_{m} \hat{\bu}_m (r) e^{i m \phi}~. 
\end{equation}
Because the fluid container is not invariant under rotation, such decomposition necessarily comprises more than one wave number, but we can still identify the largest contributions to the azimuthal wave-number decomposition by computing the energy of the coefficients by radial integration, that is: 
\begin{equation}
\label{eq:azimuthal_mode_dec_energy}
\int_{r<r_{\mathrm{max}}}  \vert \hat{\bu}_m (r)\vert^2 r \mathrm{d} r ~.
\end{equation}
where $r_{\mathrm{max}}$ is a maximum radius over which the integration is performed.
We choose $r_{\mathrm{max}} = 0.5 a$, which approximately corresponds to the radius of the largest circle fitting inside the PIV field. 

Note that no coordinate transformation correcting for the elliptical distortion is applied.
Such a transformation is not required as it complicates the wave number decomposition of smaller scales modes such as the one at $\omega = 0.61$ (see figure \ref{fig:triadic_resonance}.g), and leads to qualitatively similar decomposition for larger scales modes such as those shown in figures \ref{fig:triadic_resonance} (c to f).
Calling $m_i$  the wavenumber of the resonant mode (here $\pm 1$, see figure \ref{fig:triadic_resonance}b) and $m_{j,k}$ the daughter waves', significant non-linear interaction is ensured provided that:
\begin{equation}
\label{eq:resoance_wave_number}
 m_j + m_k = m_i
\end{equation}
which is very similar to the condition on frequency (\ref{eq:triadic_resonance_condition}). 
In the present case, the mode $i$ is the resonant one, for which $m_i = \pm 1$, so that the difference between absolute value of the main wave numbers of paired modes should be $\pm 1$. 
The result of computing the azimuthal wave number decomposition is shown in figure \ref{fig:azimuthal_wave_number}, and is, at least for two pairs of modes, consistent with the relation (\ref{eq:resoance_wave_number}).  
Lastly, regardless of considerations on the azimuthal structures, it is striking in figure \ref{fig:triadic_resonance} that only modes with similar scales are observed to couple in triadic resonances. 
This is coherent with the fact that, as the resonant mode at $\omega = 2.00$ has almost no horizontal variations, the two daughter modes must have matching horizontal wave numbers to ensure a significant spatial overlap and efficient energy transfer.
To conclude, the spatio-temporal analysis of the flow at low forcing amplitudes provides significant evidence that, in this regime, the non-linear saturation of the elliptical instability is a superposition of daughter waves that are in triadic resonant interaction with the unstable mode.
This state is robust as the Ekman number is changed, provided the forcing remains small.
It could be seen as discrete inertial wave turbulence \citep{kartashova_discrete_2009-1}, which should lead to inertial wave turbulence in the low dissipation and forcing regime, as shown by \cite{le_reun_inertial_2017-1} with an idealised numerical model. 
To finish with, one could argue that in the present
set-up, some hypotheses of classical wave turbulence theory \citep{ nazarenko_wave_2011} are not satisfied. 
First, the flow is not statistically
homogeneous in space, as we observe modes whose structures are
influenced by boundaries. 
It is neither statistically homogeneous in
time because the system is by essence forced at a single frequency.
Lastly, as discussed above, dissipation is still too high and prevents
energy cascade in a large range of scales. 
Some of these limitations
are inherent to the experimental approach used here, but may be
irrelevant in the asymptotic geophysical regime. 
Note also that none
of the previous numerical or experimental set-ups studying rotating
turbulence have yet reported a regime dominated by inertial waves only and without a dominant geostrophic component (see
\cite{godeferd_structure_2015} for a review) as we observe here. 
We thus
argue that this work is a first step towards reaching an inertial wave
turbulence regime.

\section{Large forcing amplitudes: a geostrophic-dominated regime}
In this section, we detail our experimental finding that the wave-dominated regime vanishes at larger forcing amplitude.
This secondary transition is caused by the emergence of a strong geostrophic anticyclonic vortex that back-reacts on the structure of the waves.
Our experimental set-up allows to locate the secondary transition in the forcing-dissipation $(Ro_i, E)$ plane and to find the existence domain of the wave-dominated regime. 
As earlier, our analysis starts by exploring the spectral content of the saturation flow. 
We then detail its shape and amplitude and how it affects the resonant modes.

\subsection{Spectral content of the saturation flow at large forcing amplitude}

\begin{figure}
\centering
\includegraphics[width=\linewidth]{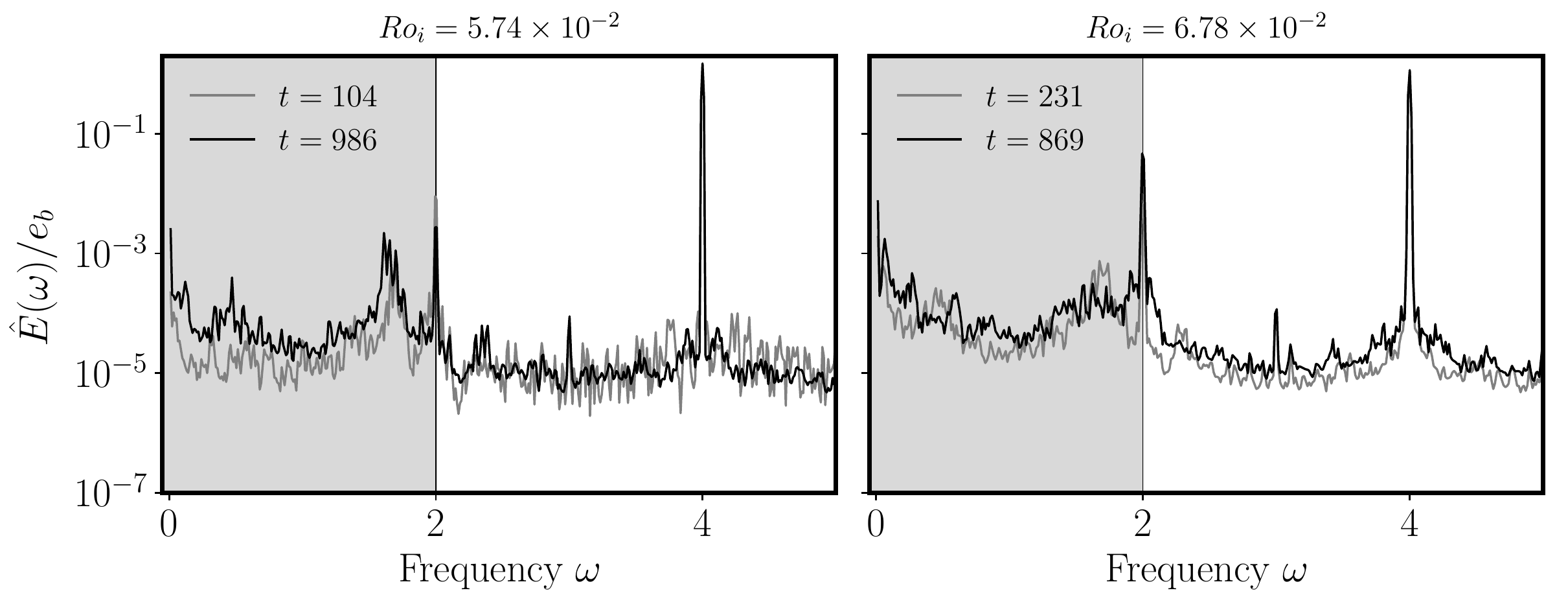}
\caption{Temporal spectra of the saturation flow obtained at $E = 5.0 \times 10^{-6}$ with the same process as in figure \ref{fig:O30RPM_low_forcing}, but with higher forcing amplitudes or $Ro_i$ values. 
As earlier, the inertial wave domain is highlighted in grey, and the base flow and resonant peaks are clearly identifiable. 
}
\label{fig:O30RPM_power_spectrum}
\end{figure}

We reproduce the temporal analysis of the saturation flow detailed in paragraph \ref{sec:wave_dominated_spectral_content}, now applied to the larger forcing experiments.
Several striking changes appear in these power spectra  displayed in figure \ref{fig:O30RPM_power_spectrum} compared to what has been obtained at lower forcing (see figure \ref{fig:O30RPM_low_forcing}).
First, the number of peaks in the inertial wave domain is reduced, and the remaining ones are wider. 
Moreover, the ratio between the background level and  the forcing peak (at $\omega = f = 4$) increases with the forcing, thus suggesting that the flow becomes more turbulent. 
Lastly, the gap between the mean flow (at $\omega  = 0$) and the resonant mode reduces from two to less than one order of magnitude between low (figure \ref{fig:O30RPM_low_forcing}) and large forcing.
It is thus clear that the saturation flow has transitioned towards another regime: the many triadic resonances that clearly appeared in the inertial modes domain in figures \ref{fig:O30RPM_low_forcing} and \ref{fig:triadic_resonance} are no longer present.

\subsection{The emergence of a strong geostrophic anticyclonic vortex}

\label{sec:geostrophic_emergence}

\begin{figure}
\includegraphics[width=\linewidth]{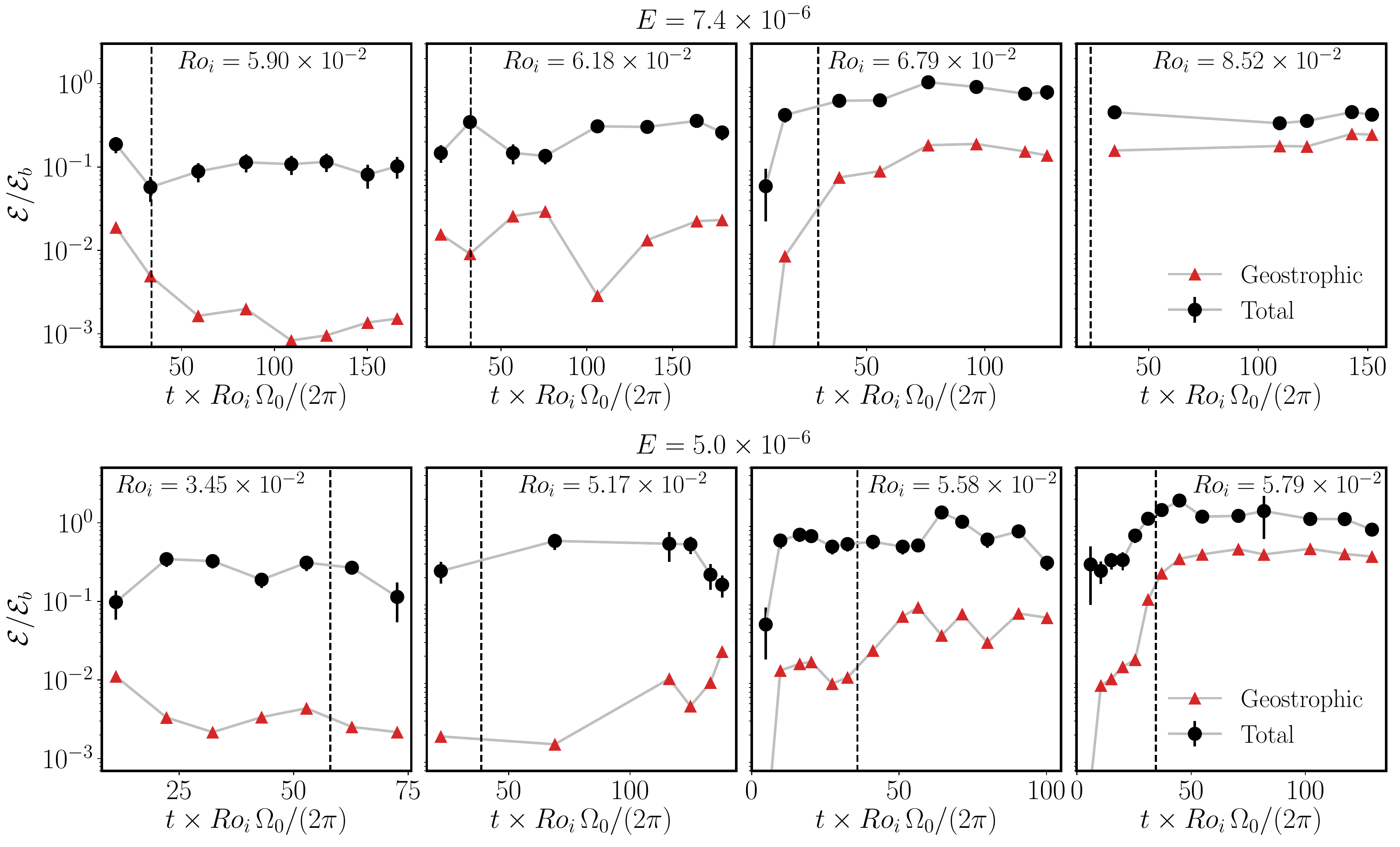}
\caption{
Long-term evolution of the kinetic energy of the total perturbation flow $\bu$ (see section \ref{sec:kinetic_energy}) and the geostrophic component for experiments carried out at $E = 7.4 \times 10^{-6}$ (top panels) and $E = 5.0 \times 10^{-6}$ (bottom panels).
Each experimental run consists in a succession of recordings from which the average total kinetic energy and the mean flow kinetic energy are extracted. 
They are normalised by the base flow energy $\mathcal{E}_b$ defined in (\ref{eq:Eb_definition}). 
Note that the spectra shown in figures \ref{fig:O30RPM_low_forcing} and \ref{fig:triadic_resonance}-a correspond to the time series $E = 5.0 \times 10^{-6}$, $Ro_i = 5.17 \times 10^{-2}$ and $E = 7.4 \times 10^{-6}$, $Ro_i = 5.90 \times 10^{-2}$.
The errorbars on the total kinetic energy correspond to its standard deviation over each recording.
For each experiment, time is normalised by a non-linear timescale $ 2 \pi / (\Omega_0 Ro_i)$. 
The vertical dashed line denotes a long non-linear timescale $2 \pi / (\Omega_0 Ro_i^2)$. 
}
\label{fig:kinetic_geostrophic_timeseries} 
\end{figure}

The key to the transition out of the wave-dominated regime is the rise of the geostrophic mean flow amplitude, a feature noticed in figure \ref{fig:O30RPM_power_spectrum}.
The aim of the present paragraph is to explain the properties of this particular component of the saturation flow and to quantify its evolution when the forcing amplitude, or equivalently $Ro_i$, is changed.

The geostrophic component of the flow $\overline{\bU}$ is extracted by time averaging the velocity fields over a sufficiently long time period.
It is found that 5 rotation periods are sufficient to average a representative mean flow, although we perform the averaging operation over the full length of a recording, which typically ranges from tens to about a hundred rotation periods.
As an experimental run comprises several successive recordings, we can characterise the long-term evolution of the geostrophic component of the flow. 
We show in figure \ref{fig:kinetic_geostrophic_timeseries} time series of the kinetic energy of the geostrophic component extracted from the saturation flow for increasing forcing amplitude or $Ro_i$.
The average over each recording of kinetic energy of the total perturbation flow $\bu$, computed in section \ref{sec:kinetic_energy}, is also given for reference. 
To facilitate the comparison between experiments with different forcing amplitudes $Ro_i$, time is normalised by the non-linear time scale $2 \pi/(\Omega_0 Ro_i)$.
For the majority of the experiments, both the total and mean flow kinetic energy time series reach a statistically steady state within a few tens of non-linear timescales.
Furthermore, we notice a gradual and clear change in the energy of the mean flow as $Ro_i$ is increased.
At low forcing amplitude, the geostrophic kinetic energy remains about two orders of magnitude smaller than the base flow kinetic energy.
A quite rapid transition is observed between $Ro_i = 5.90 \times 10^{-2}$ and $Ro_i = 6.79 \times 10^{-2}$ at $E = 7.4 \times 10^{-6}$, and between $Ro_i = 5.17 \times 10^{-2}$ and $Ro_i = 5.79 \times 10^{-2}$ at $E = 5.0 \times 10^{-6}$, where the geostrophic kinetic energy increases by two orders of magnitude.  
\begin{figure}
\includegraphics[width=\linewidth]{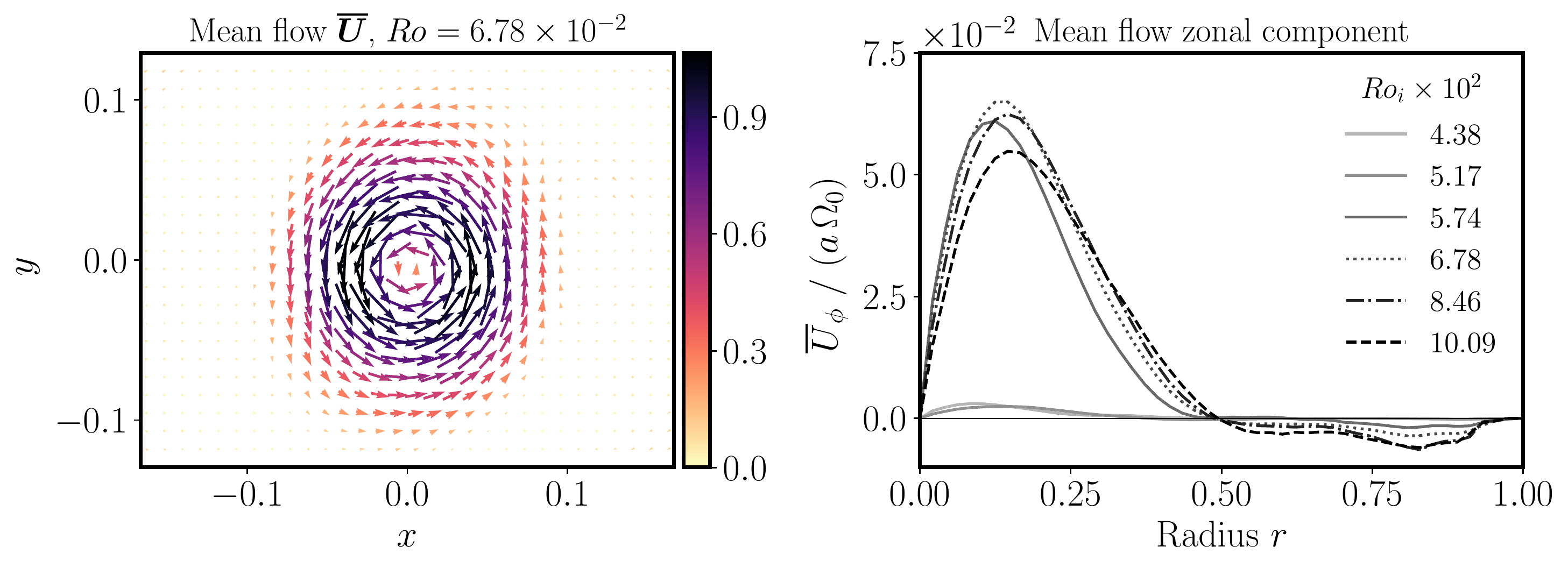}
\caption{\textbf{Left}: mean flow $\overline{\bU}$ extracted by averaging of the saturation flow observed for $f= 4$ at an Ekman number $E = 5.0 \times 10^{-6}$ and an input Rossby number $Ro_i = 6.78 \times 10^{-2}$. The colorscale of the arrows give the norm of the velocity, normalised by the base flow velocity $Ro_i \Omega a$. 
The rotation of the turntable being clockwise, the observed vortex is an anticyclone. 
Lengths are normalised by $a$.
\textbf{Right}: zonal average of the mean flow orthoradial velocity at the same Ekman number;
it is normalised by $a \Omega_0$, independently of $Ro_i$, which enhances the sharp change in the mean flow amplitude. 
  }
\label{fig:mean_flow}
\end{figure}

Regardless of the forcing amplitude and the rotation rate, the mean flow $\overline{\bU}$ always adopts the shape of an anticyclonic zonal wind, \ie a stationary flow with mostly azimuthal velocity and which rotates counter-clockwise (while the rotation of the turntable is clockwise).
A typical view of the mean flow is given in the left panel of figure \ref{fig:mean_flow} for a value of the forcing amplitude $Ro_i$ where the geostrophic energy reaches a high value.
To better quantify the structure and the amplitude of the meanflow $\MF$, we consider its zonal average $\overline{U}_\phi (r)$, defined as :
\begin{equation}
\overline{U}_\phi (r) = \frac{1}{2\pi} \int_{\phi = 0}^{2 \pi} \MF \cdot \mbf{e}_\phi ~ \mathrm{d} \phi ~.
\end{equation}
It is computed by dividing the PIV field in concentric rings of radius $r$ and averaging the quantity $\MF \cdot \mbf{e}_\phi$ on each ring. 
This zonal average is pictured in figure \ref{fig:mean_flow} right.
We retrieve a clear transition from the lowest values of the input Rossby number ($Ro_{i} < 5.28 \times 10^{-3}$ in figure \ref{fig:mean_flow}-right) with very weak vortex amplitude to larger values where a strong mean flow develops.
A similar strong anticyclonic vortex was also observed by \cite{favier_generation_2015} and \cite{grannan_experimental_2014} in the turbulent saturation of the libration-driven elliptical instability, at Ekman numbers ranging from $5 \times 10^{-5}$ to $10^{-4}$ and at a single input Rossby number $Ro_i = 0.272$. 
\begin{figure}
\centering
\includegraphics[width=\linewidth]{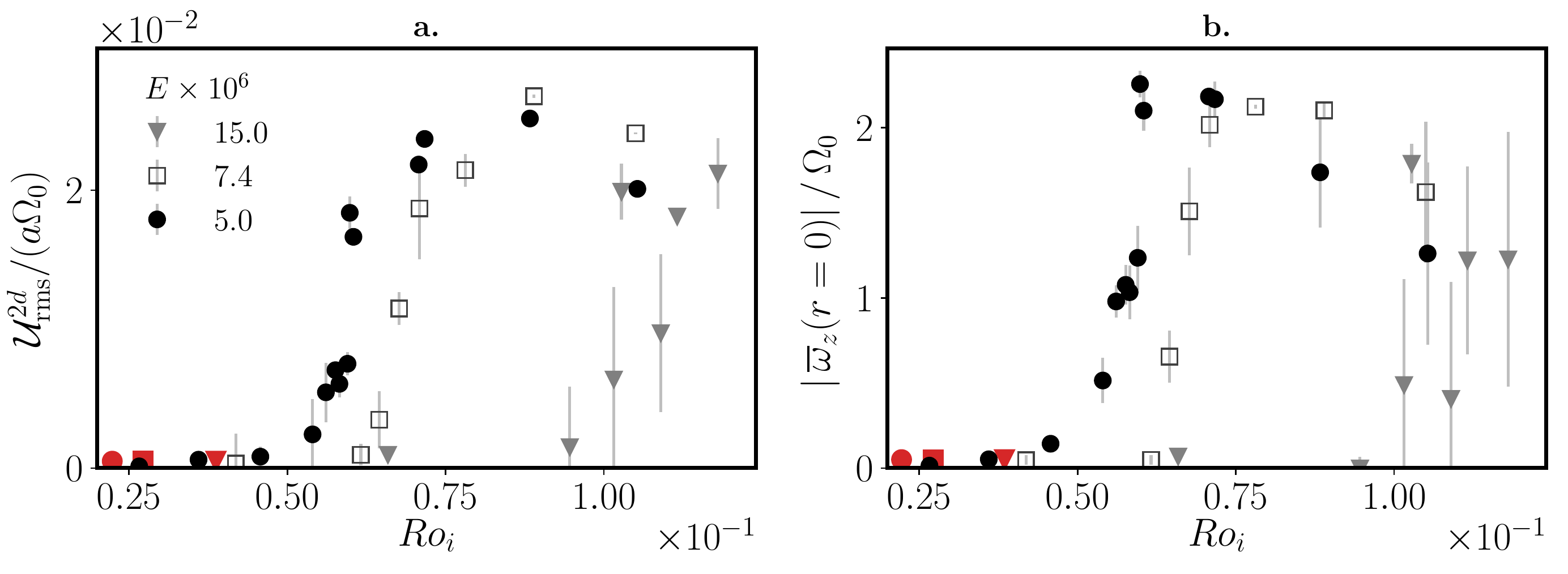}
\caption{ \textbf{a}: experimental measurements of the root mean square value $\mathcal{U}_{\mathrm{rms}}^{2d}$ of the mean flow, defined in equation (\ref{eq:rms_mean_flow}), as a function of the input Rossby number $Ro_i$; it is normalised by $a \Omega_0$ instead of $Ro_i a \Omega_0$ so that the amplitude of the rms value is independent of $Ro_i$.
The red symbols materialise the approximate critical Rossby number  $10 \sqrt{E}$ at the Ekman number corresponding to the shape of the symbol. 
\textbf{b}: experimental measurements of the central vorticity of the mean flow  $\overline{\omega}_z (r = 0)$ (see formula (\ref{eq:zonal_vorticity})) as a function of $Ro_i$. It is normalised by the rotation rate $\Omega_0$. 
Note the the mean flow is an anticyclonic vortex and has therefore a vertical vorticity opposite to the rotation.
For both panels, the error bars indicate the typical variability in the saturation phase by computing the standard deviation of $\mathcal{U}_{\mathrm{rms}}^{2d}$ and $\overline{\omega}_z (r= 0)$ when more than one data set was available.
}
\label{fig:30RPM_geostrophic_energy}
\end{figure}
To further quantify the transition, we introduce two diagnostic quantities.
The first one is the root mean square value of the mean flow $\mathcal{U}_{\mathrm{rms}}^{2d}$ defined as:
\begin{equation}
\label{eq:rms_mean_flow}
(\mathcal{U}_{\mathrm{rms}}^{2d})^2 ~=~ \left\langle \overline{\bU}^2 \right\rangle
\end{equation}
where the symbol $\left\langle \, \cdot \, \right\rangle$ denotes averaging over the whole PIV field, which we recall is in the equatorial plane of the ellipsoid.
The second one is the central vorticity of the mean flow to quantify the central rotation rate of the geostrophic vortex. 
The average radial vorticity profiles $\overline{\omega}_z (r)$ is computed as: 
\begin{equation}
\label{eq:zonal_vorticity}
\overline{\omega}_z (r) = \frac{\overline{U}_\phi}{r} + \der{\overline{U}_\phi}{r}~.
\end{equation}
Close to the centre, it is computed via fitting a 3rd order polynomial to the zonal velocity profile $\overline{U}_\phi$ as those depicted in figure \ref{fig:mean_flow};
$\overline{\omega}_z (r)$ is then twice the first order coefficient of the polynomial fit. 
For each experiment, these two diagnostic quantities are averaged over the last half of the recordings, for which a statistically steady state is usually obtained (see figure \ref{fig:kinetic_geostrophic_timeseries}). 
The evolution of the  mean flow rms and central vorticity is shown in figure \ref{fig:30RPM_geostrophic_energy}. 
In both panels, a clear transition from a negligible amplitude mean flow to a strong mean flow is observed, with a critical Rossby number $Ro_{i,c}$ depending on the Ekman number. 
It proves the existence of a secondary transition of the geostrophic anticyclone that builds on the turbulent saturation of the libration-driven elliptical instability.
%
%

%
Below the secondary transition, the variations of the central vorticity with $Ro_i$ do not match any scaling $\overline{\omega}_z (r) \propto Ro_i^2$ that has been proposed and measured for mean flows driven by the libration base flow alone \citep{busse_mean_2010,sauret_experimental_2010,sauret_libration-induced_2013} or by non-linear self interaction of a single inertial mode in the laminar regime \citep{tilgner_zonal_2007-1,sauret_tide-driven_2014,morize_experimental_2010-1}.
This shows the mean flow does not result from non-linear self -interaction in the boundary layer of simple structures such as the base flow or a single inertial mode. 

%
For input Rossby numbers above the secondary transition, the central vorticity reaches a plateau at the value of $\sim - 2 \Omega_0$, meaning that the core of the vortex cancels out on average the rotation of the fluid. 
It is striking that the central vorticity saturates at this value for which anticyclones are marginally stable according to the Rayleigh criterion for centrifugal instability \citep{drazin_hydrodynamic_2004,afanasyev_three-dimensional_1998}.
Such a saturation value emerging out of a turbulent saturation may be reminiscent of self-organised criticality \citep{bak_self-organized_1987}, an idea that is for instance invoked in stratified turbulence where flows are thought to be maintained close to marginal stability respective to shear instability (see for instance \cite{salehipour_self-organized_2018}).
Beyond this plateau, a slight decrease of both the rms and the central vorticity of the mean flow is observed at higher $Ro_i$, at least for Ekman numbers $E = 5.0 \times 10^{-6}$ and $E = 7.4 \times 10^{-6}$.
It is possibly linked to a transition from rotating to isotropic turbulence as the Rossby number draws closer to $1$ \citep{yokoyama_hysteretic_2017}.
As observed for instance by \cite{barker_non-linear_2013-1}, increasing the amplitude of the forcing leads to the destabilisation of the strong vortices produced by the saturation of the elliptical instability. 
Nevertheless, in our set-up, the transition from a purely two-dimensional vortex to three-dimensional structures with possible vertical motion is difficult to further quantify.  
%


%

%
\subsection{Building a regime diagram of the saturation flow}

\begin{figure}
\centering
\includegraphics[width=0.55\linewidth]{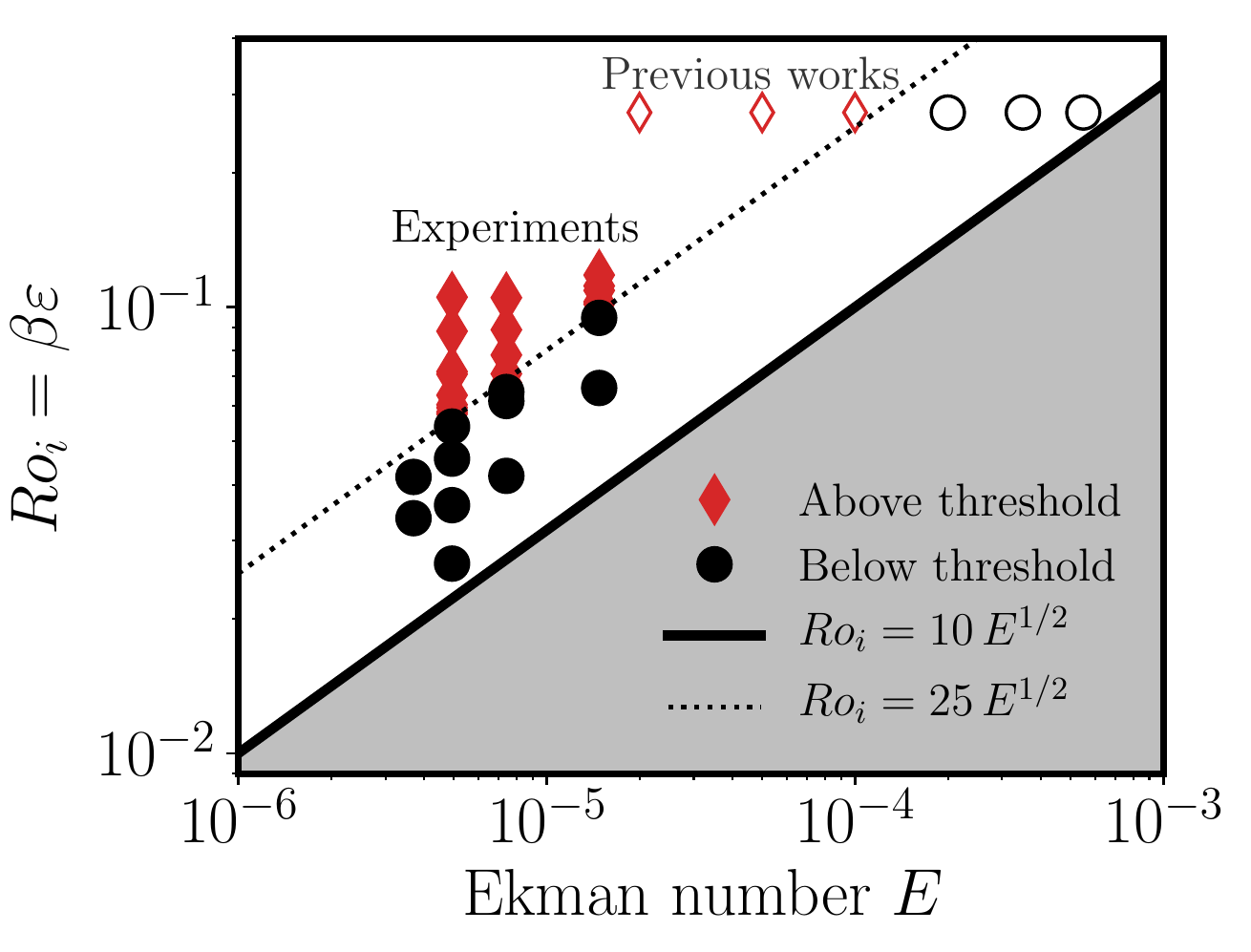}
\caption{Experimental determination of the position of the secondary instability of the geostrophic vortex as a function of the Ekman number $E$ and the input Rossby number $Ro_i$. 
The shaded area denotes the values of the control parameters for which the libration base flow is stable, its upper bound is a line $Ro_i = 10 \sqrt{E}$.
The empty symbols stand for the control parameters of \cite{favier_generation_2015} and \cite{grannan_experimental_2014} for which a central anticyclone similar to what is observed here is obtained in the non-linear saturation.
As it can be noticed in the data of \cite{favier_generation_2015} (figure 5), there exists a sharp increase in the central vorticity; points below this transition are black and points above are red.
To determine the location of the transition in our experiments, we consider that the threshold is reached when the energy of the mean flow is larger than $15$ \% of its largest value over a whole set of experiments carried out at the same Ekman number but different input Rossby numbers.
Varying this threshold, say from $5$\% to $20$ \%, may change the nature of the points around the dashed line, but does not affect the overall trend for the separation between the two regimes. 
}
\label{fig:f4_regime_diagram}
\end{figure}

As noted in figure \ref{fig:30RPM_geostrophic_energy}, the secondary transition between a wave-dominated and a geostrophic-dominated regimes is characterised by a sudden increase in the mean flow  rms velocity and vorticity.
The critical value of $Ro_i$ at which this transition occurs varies with the Ekman number.
Our experiments therefore allow to propose a regime diagram of the saturation of the instability based on the mean flow diagnostic quantities.
We show in figure \ref{fig:f4_regime_diagram} the location of all the experiments that have been carried out at a forcing frequency of $f = 4$ with different input Rossby numbers $Ro_i$ and at different rotation rates or Ekman numbers (see table \ref{tab:runs_statistical}). 
The experiments below and above the secondary transition are discriminated by setting a threshold on $\mathcal{U}_{\mathrm{rms}}^{2d}$ to $15$ \% of the maximal value over all the experiments performed at the same rotation rate. 
This regime diagram reveals that the critical value of the input Rossby number for the secondary transition $Ro_{i,c}$ follows a power law respective to the Ekman number, that is:
\begin{equation}
\label{eq:critical_scaling}
Ro_{i,c} \propto E^{1/2}
\end{equation}
although a definitive power law would require a larger range of Ekman numbers. 
We also report for comparison in figure \ref{fig:f4_regime_diagram} the location of the control parameters explored by \cite{grannan_experimental_2014} and \cite{favier_generation_2015} and the points at which they observed either a strong or weak vorticity mean flow.
The location of the secondary transition they report is consistent with ours. 
Note however that the state they observed below the secondary transition could not  be convincingly described as an inertial wave turbulence due to the larger values of the Ekman number and input Rossby number they explore.

\subsection{Back-reaction of the geostrophic vortex on the resonant modes}

\begin{figure}
\centering
\includegraphics[width=0.55\linewidth]{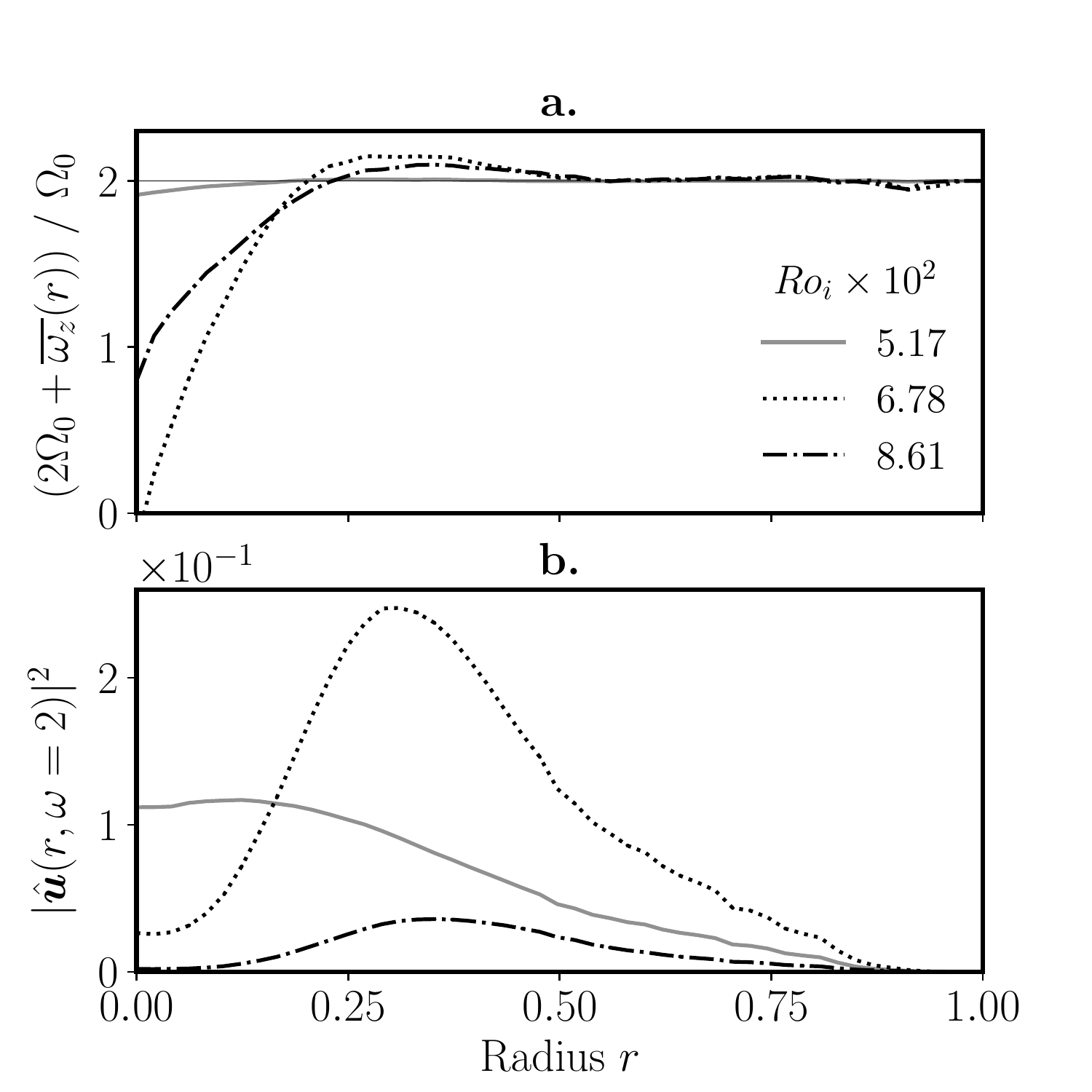}
\caption{ \textbf{a.} : total vertical vorticity $2 \Omega_0 + \overline{\omega}_z(r)$ of the fluid inside the ellipsoid as a function of the radius. \textbf{b.}: local kinetic energy of the resonant mode at $\omega = 2$ as a function of the radius. The energy is normalised by $(Ro_i a \Omega_0)^2$ and the radius by $a$. }
\label{fig:radial_resonant_energy}
\end{figure}

\begin{figure}
\includegraphics[width=\linewidth]{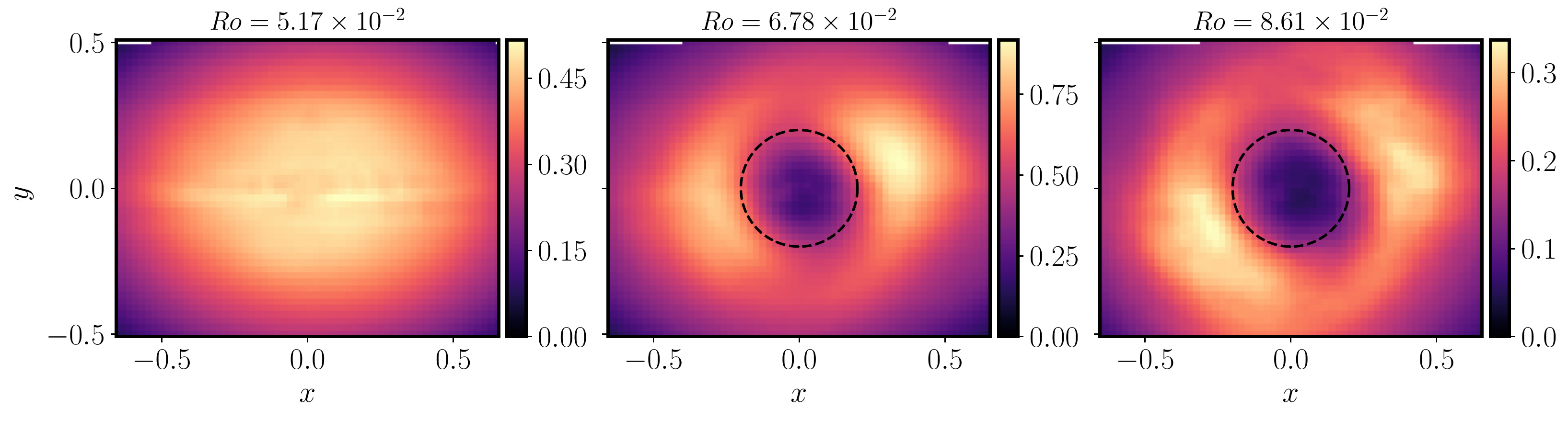}
\caption{Norm of the velocity of the resonant wave extracted by temporal filtering of the saturation field below (\textbf{left)} and above (\textbf{centre} and \textbf{right}) the threshold of the secondary instability on the geostrophic anticyclone.
 The dashed circle materialises the radius at which the global vorticity is always below $2 \Omega_0$, \ie $r/a \sim 0.2$ (see figure \ref{fig:radial_resonant_energy}). The colorscale gives the norm of the velocity normalised by $Ro_i a \Omega_0$.  }
\label{fig:resonant_wave_structure}
\end{figure}

As it may be noticed from the previous results, one of the striking consequences of the emergence of a strong geostrophic anticyclone is the loss of the numerous and well-defined triadic resonances detailed in the previous paragraph. 
The blurring of triadic resonances by geostrophic vortices is a well known feature of randomly-forced rotating turbulence: advection of inertial waves by these slowly evolving modes Doppler-shifts their frequencies and force them out of resonant interaction \citep{campagne_disentangling_2015,oks_inverse_2017}. 
Nevertheless, the present case is different from usual rotating turbulence since energy is only supplied to the system through a pair of inertial modes via the elliptical instability resonance.
%
%
The situation here is rather different since the vortex is persistent and drastically changes the local rotation rate of the fluid at the centre of the ellipsoid. 
To better quantify the reaction of the waves to the existence of the strong anticyclone, we first consider the total vertical vorticity $2 \Omega_0 + \overline{\omega}_z(r)$, which is shown in figure \ref{fig:radial_resonant_energy}, and where $\overline{\omega}_z$ is determined with equation (\ref{eq:zonal_vorticity}).
There exists a radius below which the global vorticity remains significantly below $2 \Omega_0$, at the core of the anticyclonic vortex. 
Once the secondary instability develops, the core of the anticyclone cannot sustain the resonant mode at $\omega = 2$ since the maximal frequency cannot exceed the total vertical vorticity. 
As a consequence, the radial structure of the resonant mode, which is constantly excited by the elliptical instability, changes as the secondary instability develops. 
This is illustrated in figures \ref{fig:radial_resonant_energy} and \ref{fig:resonant_wave_structure}, where it appears clearly that the central amplitude of the resonant mode drops as the anticyclone emerges. 
Moreover, the spatial area affected by the drop in amplitude corresponds to the typical size of the vortex.
To conclude, the growth of the secondary instability alters the spatial structure of the resonant modes.
The strong anticyclone that emerges significantly alters the global vorticity, and hence the structure of inertial modes.
If they still exist, triadic resonances cannot be the same compared to the low forcing case as the structure of the modes has to account for the differential rotation introduced by the geostrophic flow. 
%
%

\section{Discussion: the origin of the secondary instability}

The spontaneous excitation of geostrophic structures by waves is a long standing issue in rotating fluid studies. 
As noted by \cite{kerswell_secondary_1999-1}, geostrophic vortices are always observed to emerge from non-linear interaction of waves in spite of the theorem proved by \cite{greenspan_non-linear_1969} stating that two inertial modes cannot transfer energy to vortices via triad-type interactions, at least in the asymptotic regime of low Rossby number. 
As explained in paragraph \ref{sec:theory_saturation},  
direct and non-resonant excitation of strong zonal flows by non-linear self-interaction of inertial modes in the boundary layer have been characterised in simulations and experiments \citep{tilgner_zonal_2007-1,sauret_tide-driven_2014,morize_experimental_2010-1}, but cannot be responsible for a sharp secondary transition, as the amplitude of the mean flow should be merely proportional to $Ro_i^2$. 
To explain the observation of strong geostrophic modes in rotating fluids, an instability involving non-triadic interactions have been proposed by \cite{kerswell_secondary_1999-1}.
%
The inviscid growth rate of such instabilities is proportional to $Ro^2$ where $Ro$ is a Rossby number referring to the amplitude of the waves. 
In figure \ref{fig:kinetic_geostrophic_timeseries}, although we do not observe any clear exponential growth in the geostrophic kinetic energy time series, we notice that the typical timescale associated to this non-triadic instability (proportional to $Ro_i^{-2}$) roughly corresponds to the period over which the mean flow reaches its saturation amplitude.  
However, the location of the secondary transition in the space of parameters (shown in figure \ref{fig:f4_regime_diagram}) does not seem to match with the mechanism of  \cite{kerswell_secondary_1999-1}.
As damping in a closed container is proportional to $\sqrt{E}$, the threshold Rossby number of the instability $Ro_c$ follows a power law \citep{kerswell_secondary_1999-1}: 
\begin{equation}
\label{eq:quartet_threshold}
Ro_c \propto E^{1/4}~.
\end{equation}
Assuming that the amplitudes of the waves saturate in $Ro_i$ in the non-linear regime of the elliptical instability, the threshold we observe in figure \ref{fig:f4_regime_diagram} for the secondary transition
is inconsistent with the scaling (\ref{eq:quartet_threshold}).  
One possibility is that the dissipation of the mode is in $E$ (\ie volumic) instead of $E^{1/2}$ (parietal), a situation that has been reported for instance by \cite{lemasquerier_librationdriven_2017} but at slightly higher Ekman numbers in presence of a solid inner core.
Another plausible way to explain the observed scaling may be that the secondary transition we detail in the present article is caused by finite Rossby effects.
In particular, a mode structure under an insufficient rotation rate could be unstable to shear-driven instabilities since it contains inflection points \citep{drazin_hydrodynamic_2004}, a mechanism that has for instance been studied in the special case of inertial wave attractors by \cite{jouve_direct_2014}.
%
The growth rate of such a secondary instability would then be proportional to the amplitude of the wave, \ie to the Rossby number.
Additional theoretical studies, beyond the scope of the present paper, are needed to conclude on that matter.  

%
%
%

\section{Conclusion}

\begin{figure}
\centering
\includegraphics[width=0.7\linewidth]{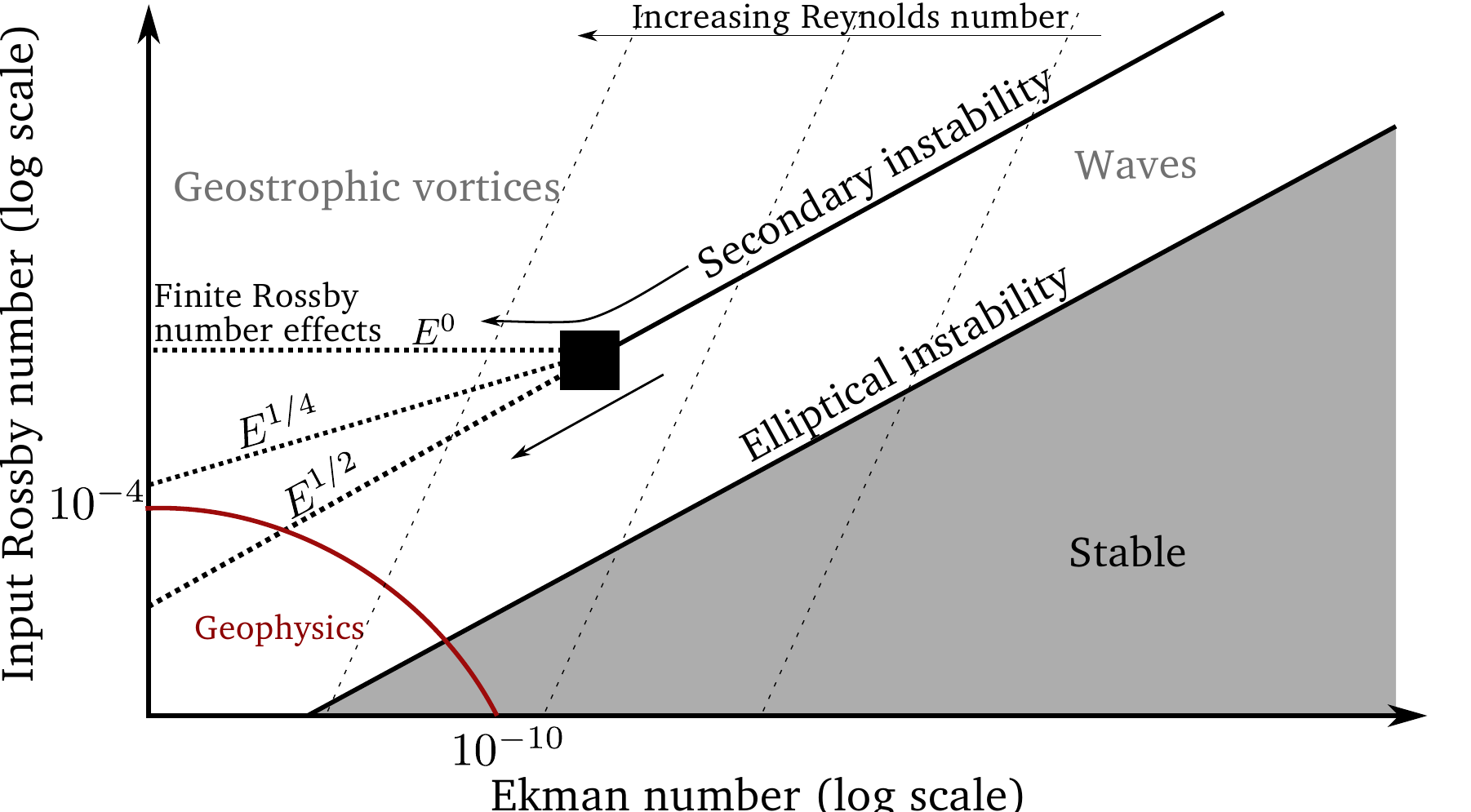}
\caption{Schematic regime diagram of the saturation of the elliptical instability proposed after the present experimental study, depending on the Ekman number $E$ and the input Rossby number $Ro_i$, building on figure \ref{fig:f4_regime_diagram}. In particular, we place the geophysically relevant regime at low $E$ and $Ro_i$. 
The elliptical instability threshold shown here is associated to a viscous damping rate dominated by wall boundary layer friction corresponding to a $Ro \propto E^{1/2}$ line. 
We report the possible behaviour of the limit between the geostrophic-dominated and the wave-dominated regimes at low $E$ and $Ro_i$ with the dotted lines: the case where the secondary instability would be a finite Rossby number effect ($E^{0}$ line), the case where the instability of \cite{kerswell_secondary_1999-1} causes the transition ($E^{1/4}$ line) and lastly the continuation of what is observed in our experiments ($E^{1/2}$ line). We also draw thin dashed line for which the Reynolds number $Re = Ro_i/ E$ is constant, and the direction in which it increases.  }
\label{fig:regime_diagram_conclusion}
\end{figure}

Throughout this paper, we have explored the non-linear fate of the libration-driven elliptical instability in low dissipation and low forcing regimes. 
Despite it being focused at one particular forcing frequency, we believe it brings some clarifications regarding the dichotomy between wave-dominated and vortex-dominated types of saturation, and their conditions of existence. 
The main result of our work is to prove the existence of a non-linear saturation regime dominated by inertial waves in triadic resonant interactions at low forcing amplitude, \ie a (discrete) inertial wave turbulence regime.
Such a regime vanishes with increasing forcing amplitudes due to a secondary transition causing the emergence of a strong geostrophic anticyclone altering the inertial modes' structure.
This transition sets the limits of the wave-dominated regime: it allows to draw a regime diagram of its existence depending on the dissipation and the forcing amplitude.
We hence find that the wave-dominated regime, which is a precursor of inertial wave turbulence, is relevant for geo- and astrophysical applications, as planets subject to the elliptical instability are usually close to the  primary instability threshold \citep{cebron_elliptical_2012}. 
When tidal interaction drives planets further away from the threshold of the instability, the expected saturation is then dominated by strong geostrophic vortices.
Our work is therefore also an experimental confirmation of the results of \cite{le_reun_inertial_2017-1} obtained with idealised numerical simulations.
It is the first time that a transition from a wave-dominated to a geostrophic-dominated regimes is clearly characterised within one experimental set-up by changing one control parameter only.

It remains to be seen whether similar transitions can be found at other frequencies, or even for other types of forcing such as those used to investigate rotating turbulence. 
Even for the present set-up, finding a frequency for which the resonant mode has a large growth rate and leads to small scale non-linear regimes, while having mostly horizontal, measurable velocities, remained challenging.
Nevertheless, the experimental work of \cite{lin_experimental_2014}, where a large-scale inertial mode at frequency $\omega = 1$ is directly forced in a precessing cylindrical annulus, suggests that the transition we find between a wave-dominated and a geostrophic-dominated regimes may be extended to other forcing frequencies.
They indeed observed at low forcing amplitude the excitation of a few short-lived triadic resonances, which vanish at larger forcing amplitudes.
Their experimental set-up did not allow to prove whether such a secondary transition is associated to the emergence of a strong mean flow, but location of the transition in the space of control parameters matches the scaling (\ref{eq:quartet_threshold}) associated to the non-triadic geostrophic instability of \cite{kerswell_secondary_1999-1}.
Although the secondary transition found in \cite{lin_experimental_2014} and in the present work may be of different origin, their observation of triadic resonant instabilities at low forcing amplitudes is consistent with our findings. 

We conclude our study by a non-linear saturation regime diagram in figure \ref{fig:regime_diagram_conclusion} based on our work, and in particular on the results of figure \ref{fig:f4_regime_diagram}.
As we cannot propose a complete description of the associated mechanism with the present data, we do not know how the secondary transition behaves as the forcing amplitude and the Ekman number are decreased towards the geophysically relevant regimes. 
In the case where the geostrophic instability is caused by finite Rossby effects, there should exist a critical input Rossby number below which only the wave type of regime is observed. 
If due to the instability mechanism \cite{kerswell_secondary_1999-1}, in the asymptotic regime of low dissipation, the secondary transition should follow a $Ro_{i,c} \propto E^{1/4}$ that is shown in the regime diagram of figure \ref{fig:regime_diagram_conclusion}.
We also report in this diagram that drawing closer to the geophysical regimes is also associated with an increase of the input Reynolds number $Ro_i/E$.
The saturation of the elliptical instability in planetary cores and in stellar interiors should therefore be more turbulent than what is observed here.
In particular, instead of the few secondary modes generated by triadic resonance that we report in this study, it is expected that a very large number of them should be excited in the form of inertial wave turbulence.
This assertion is further supported by the results of \cite{le_reun_inertial_2017-1} (supplementary material) who observed with an idealised numerical model the increase of the number of interacting inertial modes as the forcing amplitude and the Ekman number where both decreased. 
It is therefore plausible that inertial wave turbulence is excited in planetary cores and stellar interiors that are unstable to the elliptical instability. 
The consequences in terms of \textit{e.g.} energy dissipation and dynamo action remain to be seen. 

\vspace{0.5cm}

\textbf{Acknowledgement:} The authors acknowledge funding by the European Research Council under the European Union's Horizon 2020 research and innovation program through Grant No. 681835-FLUDYCO-ERC-2015-CoG. The turntable was financed by the Labex MEC (grant ANR-11-LABX-0092). The authors thank Eric Bertrand and William Le Coz for the help in the conception and construction of the set-up, as well as Jean-Jacques Lasserre for the help in tuning the PIV measurements. The authors thank Basile Gallet for insightful discussions regarding the outcomes of the present article, and J\'er\^ome Noir for helping us to implement experimental the set-up and for suggesting the use of the fluorescent particles which greatly benefited to our measurements. 
 
 \appendix
\section{Calibration of the field of view}\label{appA}

\begin{figure}
\centering
\includegraphics[width=0.9\linewidth]{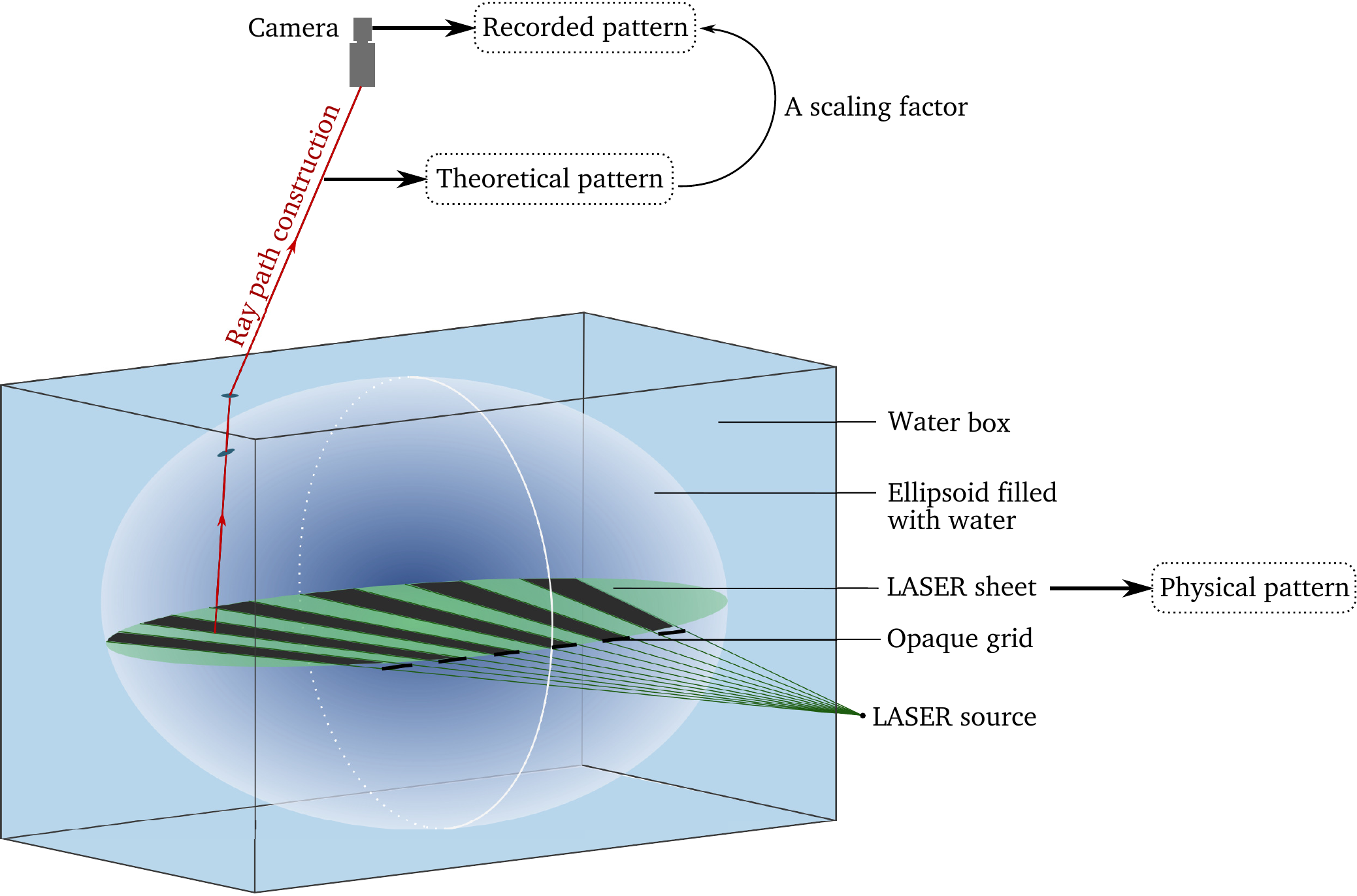}
\caption{Schematic of the set-up used to calibrate the pictures of the Laser sheet. An opaque grid pattern placed on the wall of the water box is used to interrupt the Laser rays and create a shaded pattern (the ``physical pattern'') inside the ellipsoid. The geometry of this physical pattern can be known by ray path construction. The camera records a ``recorded pattern'' that is to be compared with a ``theoretical pattern''. The latter is the result of ray path construction from the physical pattern to the camera.  }
\label{fig:calibration_setup}
\end{figure}
\begin{figure}
\centering
\includegraphics[width=\linewidth]{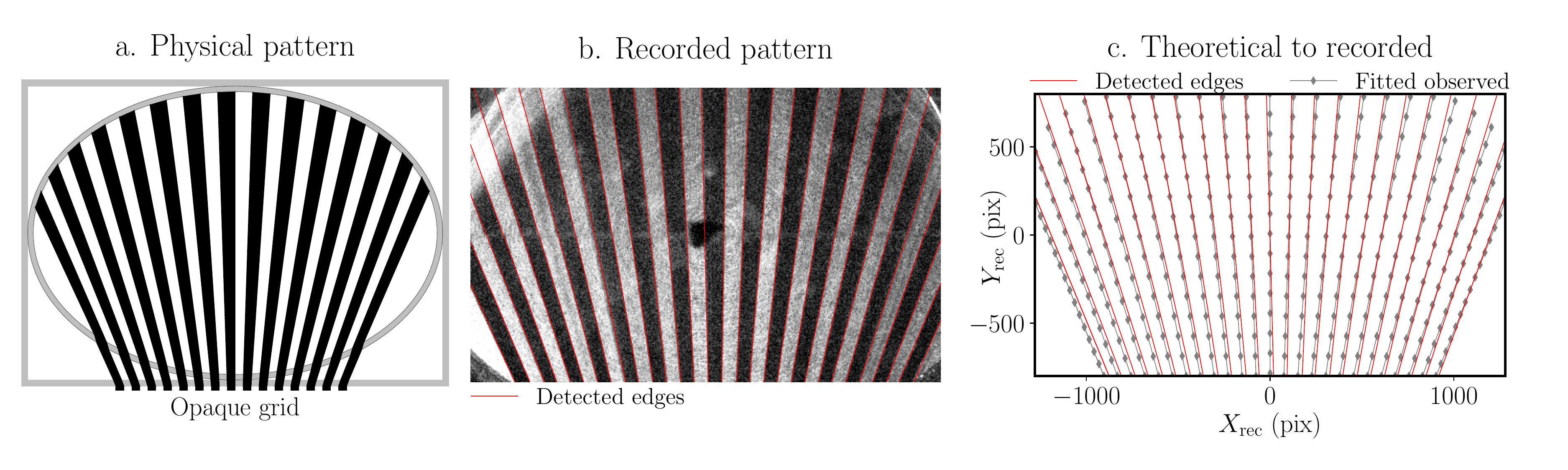}
\caption{Panels \textbf{a.} and \textbf{b.} show respectively the shaded pattern created by the an opaque grid with a $10$ mm step at the entrance of the water box. 
Panel \textbf{c.} shows the best fit between the detected edges (in red) and the theoretical pattern; the only fitting parameter is the scaling factor applied to the theoretical pattern to match the recorded pattern. 
In the computed shaded pattern depicted in \textbf{a.}, the PMMA containers, \ie the ellipsoid and the outer box, are in grey, and all these containers are filled with pure water. 
The recorded pattern of picture \textbf{b.} has been obtained by shining the Laser sheet to the particle-seeded ellipsoid while in a spin-up phase. 
It is the results of averaging 500 pictures spanning over about 15 s. 
The red lines are the result of an automated contour detection and a fit of the detected edges with lines.
Note that averaging the light diffused by particles enhances the area where the two parts of the ellipsoid are glued together, materialised by an intermediate contrast horizontal line at mid-height.
Heterogeneity in the diffused light caused by unavoidable parasitic reflections of the Laser sheet on the walls are also noticeable.    }
\label{fig:shade_pattern}
\end{figure}

As explained in section \ref{sec:experimental_methods}, the only access to the interior of the ellipsoid is a 5 mm-diameter hole. 
There is no straightforward option to perform a calibration for PIV with the use of a precise grid. 
We have therefore implemented a non-invasive method, which is less accurate, but still gives a satisfying estimate of the scaling factor between the physical field and the camera images, plus a quantitative estimate of the optical distortion on the plane of observation.
The set-up used for the calibration process is presented in figure \ref{fig:calibration_setup}. 
A shaded pattern (referenced as the ``physical pattern'') is created in the Laser sheet with an opaque grid placed on the wall of the water box. 
This grid is printed on a transparent slide and is made of black and transparent stripes, all $10.0 \pm 0.5$ mm wide.
The geometry of the physical pattern is computed from ray path construction, knowing the geometry of all the interfaces and their refractive indices; it is shown in figure \ref{fig:shade_pattern}a. 
The camera saves a ``recorded pattern'' of light and shade revealed by the particles. 
We further enhance this pattern by taking $500$ pictures as the ellipsoid spins up from 0 to 10 RPM and averaging the light pattern over this set of images. 
A typical result of this process is shown in figure \ref{fig:shade_pattern}b. 
The core of the calibration process is to produce a ``theoretical pattern'', which is the shaded pattern as it should be seen from the camera location.
It is computed from the ``physical pattern'' with ray path construction from the camera to the equatorial plane of the ellipsoid. 
The recorded and theoretical patterns are then related by a scaling factor.

The result of this process is shown in figure \ref{fig:shade_pattern} where the physical and recorded patterns are presented in panel a. and b. respectively. 
The refractive indices of water and PMMA taken for the ray path construction are respectively $ n = 1.33$ and $n = 1.49$ \citep{weber_handbook_2018}. 
The result of applying a scaling factor to the theoretical pattern to fit the recorded pattern is shown in figure \ref{fig:calibration_setup}c.
Additional effects such as camera lens distortion and local PMMA defects are within the errorbar.
The theoretical mapping between the physical and theoretical patterns also reveals that the optical distortion of the physical pattern by the water box and the ellipsoid is merely isotropic,
the relative discrepancy between the scaling in $X $ and $Y$ directions being below $2$ \%.
The fitted scaling factor between the theoretical pattern (in meters) and the recorded pattern (in pixels) is $5.69 \pm 0.08 \times 10^3$ pixels.m$^{-1}$.
The error bar accounts for the uncertainty on the respective position of the Laser sheet source point and of the ellipsoid centre, the distance between them being $42 \pm 0.5$ cm. 
In the processing of the PIV fields, we apply this scaling factor to the displacement measured in pixels by the PIV algorithm, and to the position of the points where the velocity is computed. 

%

\end{document}